\documentclass[longauth]{aa} % for a referee version
\usepackage{graphicx}
\usepackage[varg]{txfonts}
\usepackage[breaklinks=true]{hyperref}
\hypersetup{colorlinks=true,linkcolor=blue,citecolor=blue,urlcolor=blue}
\bibpunct{(}{)}{;}{a}{}{,}
\usepackage{textcomp}
\usepackage{multicol}
\usepackage{multirow}
\usepackage{longtable}
\usepackage{placeins}
\usepackage{booktabs}
\usepackage{natbib,twoopt}
\usepackage{physics}
\usepackage{tipa}
\usepackage{lscape}
\usepackage{afterpage}
\bibpunct{(}{)}{;}{a}{}{,}             
\makeatletter
  \newcommandtwoopt{\citeads}[3][][]{\href{http://adsabs.harvard.edu/abs/#3}%
    {\def\hyper@linkstart##1##2{}%
     \let\hyper@linkend\@empty\citealp[#1][#2]{#3}}}
  \newcommandtwoopt{\citepads}[3][][]{\href{http://adsabs.harvard.edu/abs/#3}%
    {\def\hyper@linkstart##1##2{}%
     \let\hyper@linkend\@empty\citep[#1][#2]{#3}}}
  \newcommandtwoopt{\citetads}[3][][]{\href{http://adsabs.harvard.edu/abs/#3}%
    {\def\hyper@linkstart##1##2{}%
     \let\hyper@linkend\@empty\citet[#1][#2]{#3}}}
  \newcommandtwoopt{\citeyearads}[3][][]%
    {\href{http://adsabs.harvard.edu/abs/#3}
    {\def\hyper@linkstart##1##2{}%
     \let\hyper@linkend\@empty\citeyear[#1][#2]{#3}}}
\makeatother

\newcommand{\lhood}{$\mathcal{L}(D|\theta)$\xspace}
\newcommand{\prior}{$P(\theta)$\xspace}
\newcommand{\posterior}{$P(\theta | D)$\xspace}
\newcommand{\objnamelong}{(470316)~2007~OC10\xspace}
\newcommand{\objname}{2007~OC10\xspace}

\begin{document}

   \title{Size and shape of the trans-Neptunian object (470316) 2007 OC10: Comparison with thermal data}

\author{
    J.M. Gómez-Limón \inst{\ref{IAA}}
    \and R. Leiva \inst{\ref{IAA}}
    \and J. L. Ortiz \inst{\ref{IAA}}
    \and N. Morales \inst{\ref{IAA}}
    \and M. Kretlow \inst{\ref{IAA}}
    \and M. Vara-Lubiano \inst{\ref{IAA}}
    \and P. Santos-Sanz \inst{\ref{IAA}}
    \and A. Álvarez-Candal \inst{\ref{IAA}}
    \and J. L. Rizos \inst{\ref{IAA}}
    \and R. Duffard \inst{\ref{IAA}}
    \and E. Fernández-Valenzuela \inst{\ref{FSI}}
    \and Y. Kilic \inst{\ref{IAA}}
    \and S. Cikota \inst{\ref{CAHA}}
    \and B. Sicardy \inst{\ref{LESIA}}
    \and F. Bragas-Ribas \inst{\ref{UTFPR}, \ref{LIneA}}
    \and M. R. Alarcon \inst{\ref{IAC}, \ref{ULL}}
    \and S. Aliş \inst{\ref{IU}, \ref{IUObs}}
    \and Z. Benkhaldoun \inst{\ref{OUKA}}
    \and A. Burdanov \inst{\ref{MIT}}
    \and J. de Wit \inst{\ref{MIT}}
    \and S. Calavia Belloc \inst{\ref{ZGZ}}
    \and J. Calvo Fernández \inst{\ref{AAVSO}}
    \and O. Canales Moreno \inst{\ref{SABA}}
    \and G. Catanzaro \inst{\ref{INAF}}
    \and S. Fişek \inst{\ref{IU}, \ref{IUObs}}
    \and A. Frasca \inst{\ref{INAF}}
    \and R. Iglesias-Marzoa \inst{\ref{CEFCA}}
    \and R. Infante-Sainz \inst{\ref{CEFCA}}
    \and A. Jiménez-Guisado \inst{\ref{OAUV}}
    \and S. Kaspi \inst{\ref{WISE}}
    \and T. Kuutma \inst{\ref{CEFCA}}
    \and D. Lafuente Aznar \inst{\ref{ZGZ}}
    \and J. Licandro \inst{\ref{IAC}, \ref{ULL}}
    \and J. L. Maestre \inst{\ref{ALBOX}}
    \and P. Martorell \inst{\ref{GU}}
    \and A. Nastasi \inst{\ref{GH}}
    \and G. Occhipinti \inst{\ref{INAF}}
    \and C. Perelló \inst{\ref{SABA}, \ref{IOTA}}
    \and C. Rinner \inst{\ref{OUKA}}
    \and B. Samper-Doménech \inst{\ref{OAUV}}
    \and A. San Segundo \inst{\ref{GUIJO}}
    \and T. Santana-Ros \inst{\ref{UNIBAR}, \ref{UNIALI}}
    \and M. Serra-Ricart \inst{\ref{IAC}, \ref{ULL}}
}

\institute{
Instituto de Astrof\'isica de Andaluc\'ia, Glorieta de la Astronom\'ia S/N, 18008 Granada, Spain \label{IAA}
\and Florida Space Institute, UCF, 12354 Research Parkway, Partnership 1 Building, Room 211, Orlando, FL, USA \label{FSI}
\and Centro Astronómico Hispano en Andalucía, Observatorio de Calar Alto, Sierra de los Filabres, 04550 Gérgal, Almeria, Spain \label{CAHA}
\and LESIA, Observatoire de Paris, Universit\'e PSL, Sorbonne Universit\'e, Universit\'e de Paris, CNRS, 92190 Meudon, France \label{LESIA}
\and Federal University of Technology-Paraná (UTFPR/PPGFA), Curitiba, PR, Brazil \label{UTFPR}
\and Laboratório Interinstitucional de e-Astronomia - LIneA - and INCT do e-Universo, Av. Pastor Martin Luther King Jr, 126 - Del Castilho, Nova América Offices, Torre 3000 / sala 817, Rio de Janeiro, RJ 20765-000, Brazil \label{LIneA}
\and Instituto de Astrof\'isica de Canarias (IAC), C/Vía L\'actea s/n, 38205 La Laguna, Tenerife, Spain \label{IAC}
\and Departamento de Astrof\'isica, Universidad de La Laguna, E-38206, La Laguna (S.C. Tenerife), Spain \label{ULL}
\and Istanbul University, Faculty of Science, Department of Astronomy and Space Sciences, 34116, Beyaz{\i}t, Istanbul, Turkey \label{IU}
\and Istanbul University Observatory Application and Research Center, 34116 Istanbul, Turkey \label{IUObs}
\and Oukaimeden Observatory, High Energy Physics and Astrophysics Laboratory, FSSM, Cadi Ayyad University, Marrakech, Morocco \label{OUKA}
\and Department of Earth, Atmospheric and Planetary Science, Massachusetts Institute of Technology, Cambridge, MA, USA \label{MIT}
\and Grupo de Ocultaciones de Zaragoza \label{ZGZ}
\and The American Association of Variable Star Observers (AAVSO), 185 Alewife Brook Parkway, Suite 410, Cambridge, MA 02138, USA \label{AAVSO}
\and Agrupaci\'on Astron\'omica de Sabadell, Prat de la Riba sn, 08206 Sabadell, Spain \label{SABA}
\and INAF - Catania Astrophysical Observatory, via S. Sofia 78, 95123 Catania, Italy \label{INAF}
\and Centro de Estudios de F\'isica del Cosmos de Arag\'on, Plaza San Juan 1, 44001 Teruel, Spain \label{CEFCA}
\and Observatorio Astronómico de la Universidad de Valencia, E-46980 Paterna, Valencia, Spain \label{OAUV}
\and School of Physics and Astronomy and Wise Observatory, Tel-Aviv University, Tel-Aviv 6997801, Israel \label{WISE}
\and Observatorio de Albox, Almeria, Spain \label{ALBOX}
\and Observatorio Astronómico de Guirguillano, Plaza Consistorial 6, 31291 Guirguillano, Spain \label{GU}
\and GAL Hassin - Centro Internazionale per le Scienze Astronomiche, Via della Fontana Mitri, 90010 Isnello, Palermo, Italy \label{GH}
\and International Occultation Timing Association - European Section (IOTA/ES), Am Brombeerhag 13, 30459 Hannover, Germany \label{IOTA}
\and Observatorio El Guijo (MPC J27), Galapagar, Madrid, Spain. \label{GUIJO}
\and Institut de Ci\`{e}ncies del Cosmos (ICCUB), Universitat de Barcelona (IEEC-UB),
Carrer de Mart\'{\i} i Franqu\`{e}s, 1, 08028 Barcelona, Spain \label{UNIBAR}
\and Departamento de Fisica, Ingeniería de Sistemas y Teoría de la Señal, Universidad de Alicante, Carr. de San Vicente del Raspeig, s/n, 03690 San Vicente del Raspeig, Alicante, Spain \label{UNIALI}
}

\date{Received ; accepted}

  \abstract
  {The shapes of only 12 trans-Neptunian objects have been directly measured, offering crucial insights into their internal structure. These properties are strongly connected to the processes that shaped the early Solar System, and provide important clues about its evolution.}
   {The aim of the present work is to characterise the size, shape, geometric albedo, and beaming parameter of the TNO \objnamelong . We compared these values to the effective diameter and geometric albedo obtained from thermal data by the TNOs are Cool survey. We also combined occultation and thermal data to constrain the size of a putative unresolved satellite.}
   {We predicted an occultation of the star Gaia DR3 2727866328215869952 by \objname on 2022 August 22. Four stations detected the occultation. We implemented an elliptical shape model for the projection of \objname . Following a Bayesian approach, we obtained the posterior probability density in the model parameter space using a Markov chain Monte Carlo method. }
   {The elliptical limb of \objname has semi-axes of $ 215^{+10}_{-7} \times 141 ^{+24}_{-23}$ km, and thus the projected axis ratio is $b/a = 0.58^{+0.16}_{-0.16}$. The area-equivalent diameter is 330$^{+56}_{-55}$\,km. From our own absolute magnitude value of $H_V = 5.40 \pm 0.02$, the geometric albedo is $p_V = 11.2 ^{+2.1}_{-5.0}$ \%. Combining the occultation results with thermal data, we constrain the beaming parameter to $\eta = 1.42^{+0.75}_{-0.58}$. Occultation data reveal that the star is double. The secondary star has a position angle with respect to the primary of $56^{+3}_{-17}$ degrees, has an angular separation of $57^{+4}_{-11}$ mas, and is 1.18$^{+0.07}_{-0.07}$ magnitudes fainter than the primary.}
  {}

   \keywords{Occultations - Kuiper belt objects: individual: 2007 OC10 - Kuiper belt: general - astrometry
            }

   \maketitle
%
%-------------------------------------------------------------------

\section{Introduction}

Trans-Neptunian Objects (TNOs) are remnants of the original protoplanetary disk that orbit beyond Neptune. Although some of the largest TNOs (such as Pluto or Eris) might have experienced significant evolutionary changes, many smaller objects are believed to have remained largely unaltered, offering valuable information about the composition and conditions of the early Solar System \citep{nesvorny_dynamical_2018}.

Studying TNO shapes gives us insight into 
their bulk composition. It is generally believed that there is a material-dependent critical size above which TNOs behave as fluids and adopt hydrostatic equilibrium shapes \citep{tancredi_which_2008}. This has not been thoroughly tested for TNOs. Moreover, a precise determination of the shape can reveal the internal structure and determine ice-rock ratios \citep{rambaux_equilibrium_2017}.

Surface properties can also be used as a proxy for TNO composition \citep{barucci_chapter_2020}. Two relevant material-dependent parameters are the geometric albedo $p_V$ and the beaming parameter $\eta$. The former measures the surface reflecting properties \citep{karttunen_solar_2017}; the latter is known to be dependent on surface roughness and thermal inertia effects \citep{lellouch_tnos_2013}. However, these properties are hard to constrain if the size of the object is unknown.

A stellar occultation occurs when a TNO passes in front of a distant star, thus blocking its light temporarily. With observers in different locations, a stellar occultation can put strong constraints on the size and shape of the occulting TNO. The accuracy of this technique can be of the order of kilometres \citep{ortiz_chapter_2020}, only paralleled by in situ space missions. 

Occultations enable the computation of the geometric albedo $p_V$ when combined with independent measurements of the absolute magnitude $H_V$, as in \cite{santos-sanz_2017_2021, kretlow_physical_2024}. Occultations also allow us to determine the apparent (projected) axis ratio. This technique has revealed that some TNOs exhibit shapes that deviate from the three-dimensional hydrostatic equilibrium expected for objects with a uniform composition \citep{ortiz_size_2017, vara-lubiano_multichord_2022}. The projected axis ratio has only been tabulated for a handful of TNOs (see detailed information in Sect. \ref{sec:ShapeResult}). Acquiring a statistically significant sample of projected axis ratios is a first step if we want to infer what the three-dimensional shape distribution is in the trans-Neptunian Belt. 

Stellar occultations have also revealed the presence of rings \citep{braga-ribas_ring_2014, ortiz_size_2017, ortiz_changing_2023}, topographic features \citep{dias-oliveira_study_2017, rommel_large_2023}, and binarity \citep{buie_size_2020, leiva_stellar_2020}, and have provided accurate astrometric positions for the object down to a few milliarcseconds \citep{rommel_stellar_2020}. In the case of double stars, occultations can constrain the relative geometry and relative brightness of the stars \citep{leiva_stellar_2020}.

The object \objnamelong is a TNO discovered at Palomar Observatory on 2007 July 22. It presents an orbital eccentricity of $e = 0.29$ and an inclination of $i = 21.7 ^ \circ$. The perihelion lies at 35.53 AU (data from the IAU Minor Planet Center). According to the classification criteria presented in \cite{gladman_nomenclature_2008} and the computations in \cite{volk_dynamical_2024}, it belongs to the dynamical class of detached TNOs: non-resonant and non-scattering objects with orbital eccentricities $e>0.24$.
We highlight that (229762) G!kún\textdoublepipe \textprimstress hòmdímà (provisional designation 2007 UK126) is the only other detached TNO with published shape analysis through stellar occultations, with an area-equivalent diameter of $D_{\text{eq}} = 638 ^{+28}_{-14}$ km \citep{benedetti-rossi_results_2016}.

The TNO \objname was a target of the TNOs are Cool programme \citep{muller_tnos_2010}, conceived to derive physical and thermal properties of TNOs with \textit{Herschel} and \textit{Spitzer} data. The fit of radiometric models to \objname data provided an area-equivalent diameter (sometimes referred to as effective diameter) of $D_{\text{eq}}=309\pm37$ km and a geometric albedo of $p_V = 12.7^{+0.4}_{-2.8}$~\%, adopting an absolute magnitude of $H_V = 5.43 \pm 0.10$ \citep{santos-sanz_tnos_2012}. Comparing occultation data with thermal measurements can provide further insights into the physical properties of the object under study. Discrepancies between thermal and occultation projected size have been proposed as a way to discover unresolved satellites \citep{ortiz_large_2020, rommel_large_2023}. In addition, occultation results can be used to improve our knowledge of the beaming parameter \citep{harris_thermal_1998}.

In this paper we present the results from the occultation produced by \objname on 2022 August 22. The occultation was observed from several sites on the Mediterranean Basin (Sect. \ref{sec:observations}). We acquired complementary photometry to constrain the absolute magnitude (Sect. \ref{sec:photometry}). We adopt a parametric model (Sect. \ref{sec:model}) and a Markov chain Monte Carlo (MCMC) technique to derive a probability distribution of position, size, and shape parameters (Sect. \ref{sec:stat_approach}). Finally, we compare our size measurement with thermal data to refine thermal parameters (Sect. \ref{sec:beaming}). Given the interest and importance of finding close satellites around TNOs that cannot be resolved with ALMA or space-based telescopes, we constrain any putative undetected satellites (Sect. \ref{sec:satellite}).

\section{Observations}
\label{sec:observations}

\subsection{Prediction of the stellar occultation}

The occultation event was predicted by the Collaborative Occultation Resources and Archive (CORA) service \citep{kretlow_cora_2024}. According to Jet Propulsion Laboratory\footnote{\url{https://ssd.jpl.nasa.gov/horizons/app.html}} (JPL) ephemeris (solution~\#~12), \objname would produce an occultation of the star identified as 2727866328215869952 in \textit{Gaia} DR3 on 2022 August 22 at 01:32:37 UTC $\pm$ 0.56 min. The upper half of Table \ref{tab:star} summarises the occulted star astrometric information from \textit{Gaia} DR3 catalogue. The lower half shows the derived astrometric position and uncertainties, propagated at the occultation epoch, used in the occultation analysis. 

\begin{figure*}[h]
\centering
\includegraphics[width=\hsize]{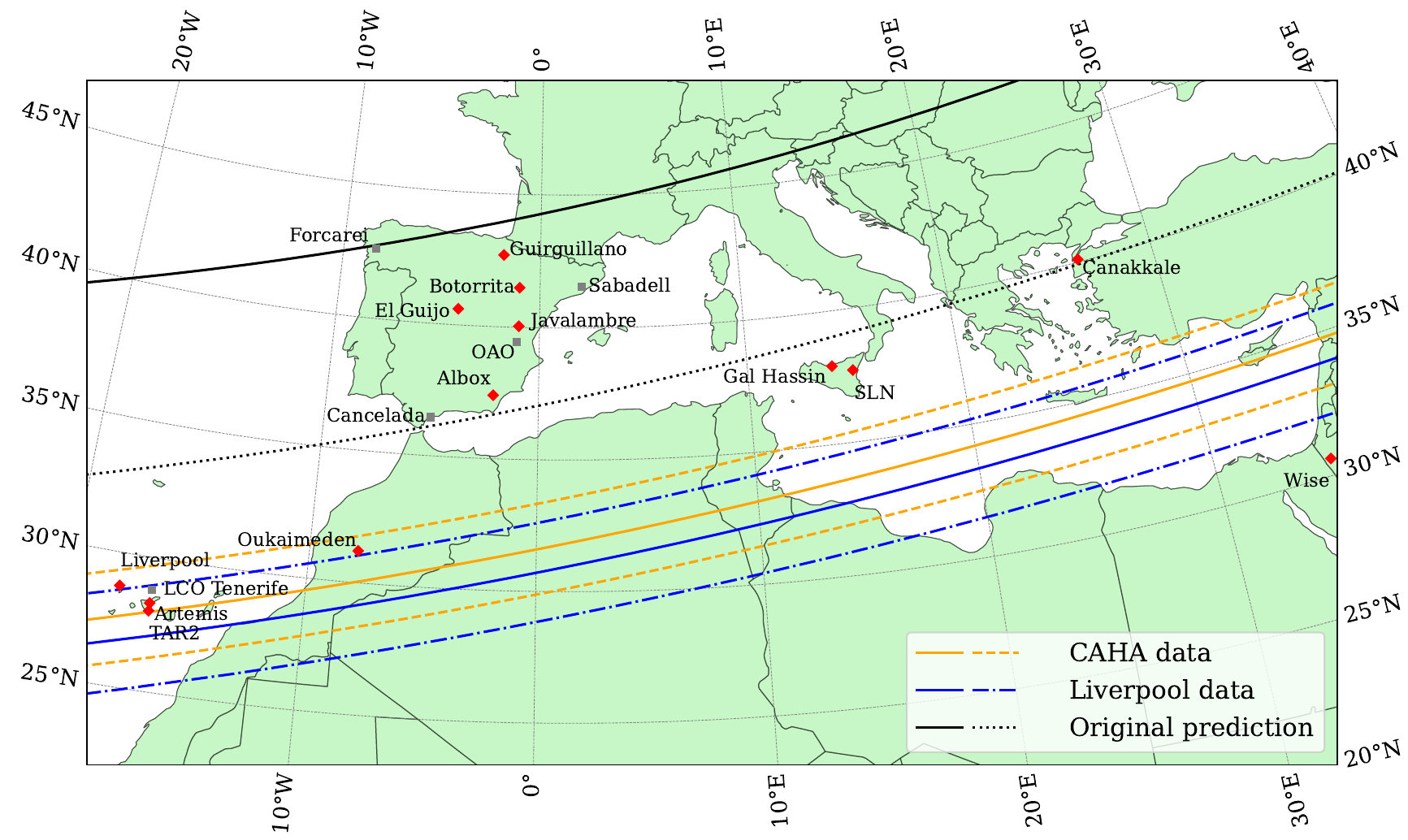}
\caption{Comparison of occultation predictions. At this time the double nature of the star was unknown. The continuous lines represent the predicted path of the centre of the shadow. The 1$\sigma$ regions along the path are represented by the broken lines. The red diamonds represent the stations that collected data. The grey squares represent the sites that suffered from bad weather or whose data were not used in the analysis. Some stations have been slightly shifted in position for improved visualisation, but the exact coordinates used in the analysis are those in Table \ref{tab:observations}.}
\label{fig:map_ini}
\end{figure*}

The original predicted shadow path, adopting a single star model, crossed continental Europe (see Fig. \ref{fig:map_ini}). It would be accessible by small-class telescopes due to the considerable brightness of the star, V = 15.23. In order to further refine this already promising occultation prediction, astrometry of \objname was acquired in the days prior to the event. The aim was to update the ephemeris of the TNO, reducing uncertainties in astrometric position. Eighteen images of \objname were taken from the Liverpool 2 m telescope in Roque de los Muchachos Observatory (La Palma, Spain) between 2022 August 5 and 2022 August 6. The 4k $\times$ 4k IO:O instrument was used, with SDSS-r filter. Sixty more images were taken from the 1.23 m telescope in Calar Alto Obervatory (Almería, Spain, abbreviated as CAHA) from 2022 August 4 to 2022 August 8. We used the 4k $\times$ 4k DLR-MKIII CCD instrument, with no filter. The exposure times ranged from 300 s to 400 s in both cases.

The average seeing on the observations from Calar Alto was 1.2 arcsec, whereas in Roque de los Muchachos it was 1.0 arcsec. The signal-to-noise ratio was, on average, of S/N=16.4 for the images of \objname taken from Calar Alto and S/N=16.6 for the case of Liverpool images.

The obtained offsets in the cross-track direction (perpendicular to the motion of \objname) from JPL \#12 ephemeris were $\Delta _{\text{Liverpool}} = -54 \pm 8$ mas and $\Delta _{\text{CAHA}} = -50 \pm 7$ mas respectively. The negative signs indicate a shift southward in the celestial sphere. The updated predictions are those shown in Fig.~\ref{fig:map_ini}.

The 1$\sigma$ region for the original prediction in Fig. \ref{fig:map_ini} is derived from the (RA, Dec) uncertainties available in JPL. For the updated predictions, they represent the formal uncertainties derived from the standard deviation of the observed astrometric points. Both updated predictions agree within these 1$\sigma$ values. It should be noted that there is a $\sim$ 60 mas difference in the observed offset in the direction of motion. This is most likely due to low accuracy in the time-stamping of the images, which is not uncommon for professional telescopes. Hence, we did not update the expected closest approach time and use the CORA reported value (extracted from JPL \#12 ephemeris).  
 
\begin{table*}[h]
\centering
\caption{Occulted star data.}
\label{tab:star}
\begin{tabular}{p{6 cm}p{7 cm}@{}}
\toprule \toprule
\multicolumn{2}{c}{\textit{Gaia} DR3 data of the occulted star, ID 2727866328215869952 } \\
\midrule
Epoch (Julian year in TCB) & 2016.0 \\
RA ($\alpha$)  & 22:08:33.066148 $\pm$ 0.05 mas\\
Dec ($\delta$)  & +12:05:55.694318 $\pm$ 0.05 mas\\
Parallax (mas) & 0.43 $\pm$ 0.06  \\
Proper motion $\alpha$ (mas yr$^{-1}$) & 4.00 $\pm$ 0.06 \\
Proper motion $\delta$ (mas yr$^{-1}$) & -2.25 $\pm$ 0.06 \\
RUWE & 1.8612541 \\
\midrule
\multicolumn{2}{c}{Astrometric coordinates propagated to the occultation epoch} \\
\midrule
Occultation epoch & 2022-08-22T01:32:37 UTC \\
$\alpha _{\text{ast}}$ & 22:08:33.067960 $\pm$ 0.42 mas \\
$\delta _{\text{ast}}$ & +12:05:55.679551 $\pm$ 0.40 mas \\
\midrule
Magnitudes \tablefootmark{a} (mag)  &  B=15.17, V=15.23, R=14.84, J=14.592 \\
 &  H=14.337, K=14.199 \\
\bottomrule
\end{tabular}
\tablefoot{The values in the first section are from \textit{Gaia} Data Release 3 \citep[DR3;][]{vallenari_gaia_2023}. The coordinates and uncertainties were propagated to the occultation epoch using \textit{Gaia} propagation tools, which take into account parallax, proper motion, and radial velocity. \tablefoottext{a}{ \footnotesize The magnitudes are from the NOMAD-1 catalogue \citep{zacharias_naval_2004}.}}

\end{table*}

\subsection{Observation of the stellar occultation}

On the 2022 August 22, 18 stations attempted an observation of the event as reported in the Tubitak Occultation Portal website \citep{kilic_occultation_2022}. Their locations are represented in Fig. \ref{fig:map_ini}. On Table \ref{tab:observations} there is detailed information about each observing site. Cancelada, Forcarei and Sabadell suffered from bad weather and no data could be collected. The rest of the stations obtained FITS images series, with the exception of Guirguillano Observatory, which recorded a video, and Las Cumbres Observatory (LCO, Tenerife), which used the drift scan technique. This technique consists in exposing for a long time without tracking, so that the stars in the field leave a trail on the image. 

From the 15 data sets obtained, 13 were used in the occultation analysis. Data from Observatorio de Aras de los Olmos (OAO) was excluded because it consisted on 300 s exposure time images, whereas the estimated maximum occultation duration was 12 s assuming a 300 km diameter. The drift scan from LCO was also discarded because of its low S/N.

Botorrita experienced synchronisation problems between the GPS and the acquisition software. The system time was stamped on the images and showed a drift of about 3 seconds with respect to the GPS. Çanakkale data also shows problems with the time stamps because these are truncated to the nearest second. However, after data reduction, it is clear that both sites did not detect the occultation (see Fig. \ref{fig:lcs}). Therefore, shifts in the timings of these data sets do not alter the result of the analysis. 

For the rest of the stations, the synchronisation of the computer time to GPS or internet NTP servers worked correctly. Their timings can be trusted to 0.1 s at least \citep[see][]{ortiz_size_2017, ortiz_large_2020}.

The data from Guirguillano station was captured at 10.24 s integrations and stored as a 25 fps video file. This is a usual technique achieved by repeating the integration into consecutive video frames. Each video frame contains timestamps from a GPS receiver, allowing the start of the integration and the exposure time during the analysis to be reconstructed. The standard processing consists of extracting the individual video frames, grouping and averaging all the frames corresponding to the same integration. The result is then saved into FITS files, storing the start of the exposure and the restored exposure time as metadata. Additional characteristics of the video data and procedure details are described in \cite{benedetti-rossi_results_2016} and \cite{buie_research_2016}.

\subsection{Data reduction: Occultation light curves}
\label{sec:data_red}
To obtain light curves for the final 13 data sets, we used our own Interactive Data Language (IDL) based routines for photometry. These routines are run on the FITS files, which have previously been dark- and flat-field calibrated, whenever calibration images were available.

These routines perform aperture photometry on the target star and on several comparison stars. Firstly, a file containing the time information is generated, with the mid-exposure time for each frame. Then, for every frame the integrated flux of the target is divided by the sum of the integrated fluxes of all comparison stars, providing a differential light curve. These fluxes are integrated using a circular aperture and are background corrected (background values computed integrating in an annulus). Finally, the resulting light curve is normalised so that the baseline level (outside any potential flux drop) is equal to one. The selection of the aperture radius was made manually for each case to maximise the light curve S/N. The same principle was applied to the selection of comparison stars.

\begin{figure*}
\centering
\includegraphics[width=0.85\hsize]{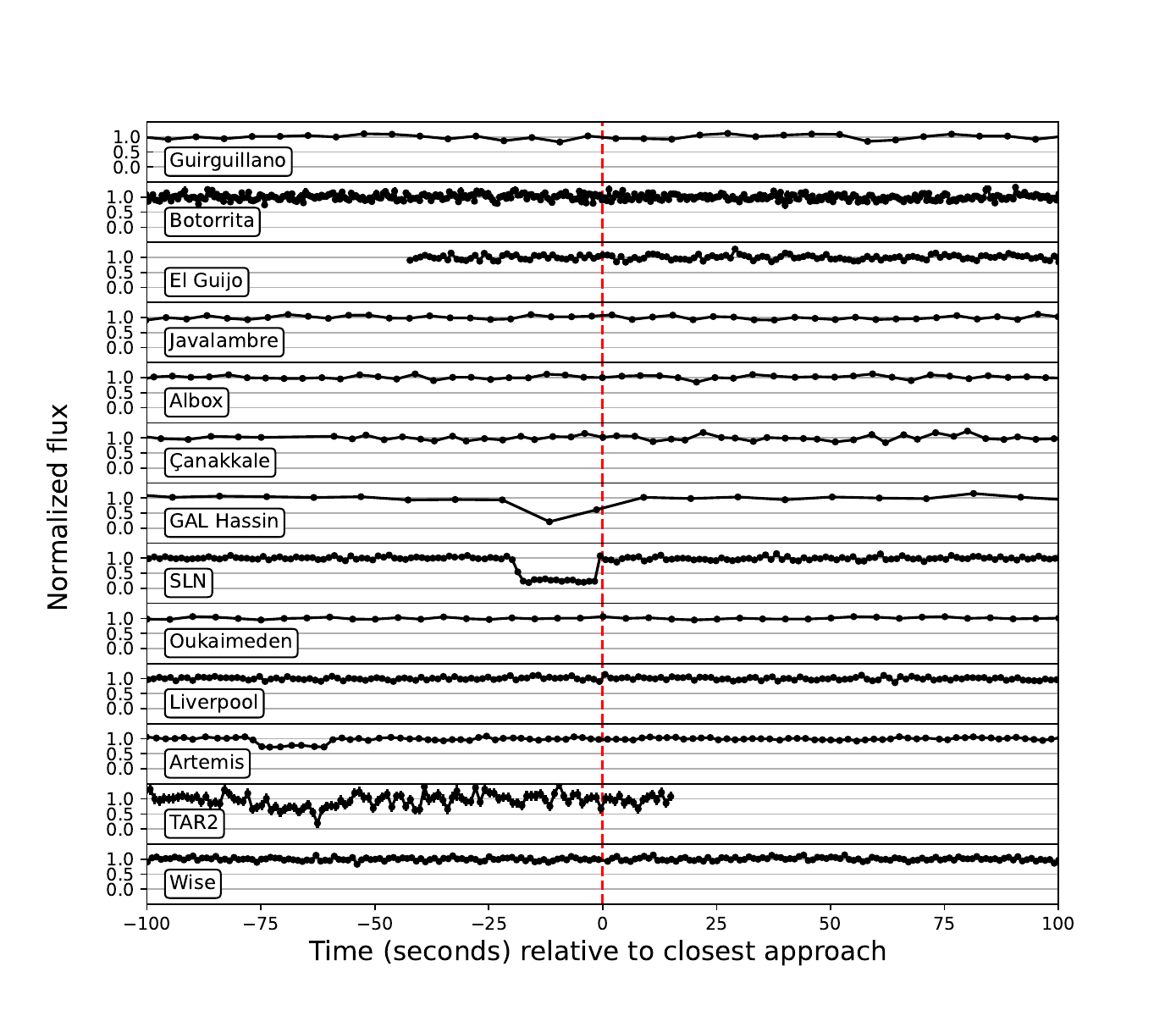}
\caption{Light curves of the stellar occultation (normalised flux vs time at mid-exposure). The origin of time is set to be the time of closest approach: 2022-08-22T01:32:37 UTC. They are ordered by distance perpendicular to the predicted shadow path (as predicted by CAHA data). The error bars are too small to be seen in the plot in most light curves, but they can be checked in the supplementary data on CDS.}
\label{fig:lcs}
\end{figure*}

The obtained light curves are plotted in Fig. \ref{fig:lcs}. They can be accessed on the CDS archive. There are clear occultation features in the curves from GAL Hassin, Serra la Nave (SLN) and Artemis, without secondary drops. In a classic approach, they would be considered positive detections. However, TAR2 light curve is much lower in S/N (Artemis aperture is 100 cm, TAR2 is 42 cm) and the flux drop is not apparent. But this telescope is only 300 m away from Artemis (both at the Teide Observatory). Therefore, it should also be a positive detection. 

A stellar flux drop of 75 \% was observed at GAL Hassin and SLN stations, whereas the drop of only 25 \% was measured at Artemis. This result was unexpected, as the predicted magnitude drop, based on the known magnitudes of the star (Table \ref{tab:star}) and the TNO (from JPL ephemeris used in CORA), was estimated to be 5.7 magnitudes, equivalent to a 99.5 \% drop in flux. Liverpool and Oukaimeden observed no flux drop, and they are located in between Sicily (GAL Hassin and SLN stations) and Tenerife (Artemis and TAR2 stations), according to the direction of the shadow path (see Fig. \ref{fig:map_ini}). These are clear indications that the target star is an optical double or a binary: The telescopes in Sicily observed the occultation of the brighter member of the pair and those in Tenerife observed the occultation of the fainter member. As seen from Liverpool and Oukaimeden, \objname passed in between the two stars, not occulting any of them. Therefore, we implemented a double star model in the analysis of the event (see Sect. \ref{sec:model}).

The fact that the star is double is coherent with the \textit{Gaia} DR3 Re-normalized Unit Weight Error value (RUWE) being $>1.4$ (see Table \ref{tab:star}). The RUWE is expected to be close to 1.0 when a single-star model fits the astrometric observations adequately. A value noticeably higher than 1.0 (e.g. >1.4) may suggest that the source is either not a single star or presents challenges for the astrometric solution \citep{castro2024gaia}.

\subsection{Photometric measurements}
\label{sec:photometry}
Photometric observations of \objname were made with the IO:O instrument of the 2m Liverpool Telescope at La Palma Observatory, the 2k$\times$2k CCD camera of the 1.5 m telescope at Sierra Nevada Observatory and the 4k$\times$4k DLR CCD camera of the 1.23 m telescope at Calar Alto observatory. Exposure times ranged from 360 s to 500 s. The observations were performed with SDSS r, Johnson Cousins R, Johnson V, and with no filters. The data reduction and analysis were the same as those described in \cite{ortiz_large_2020}.

For the absolute calibration we used the \textit{Gaia} DR3 stars in the field of view of each image and the \cite{riello_gaia_2021} transformation equations to the Johnson-Cousins system as explained in \cite{morales_absolute_2022}. The uncertainty in the absolute calibrations changed slightly from night to night with the mean value at 0.05 mag.

The observed magnitudes were translated into reduced magnitudes by subtracting $5\log _{10}(R_h \Delta)$, where $R_h$ and $\Delta$ were the heliocentric and geocentric distances of \objname. For the absolute magnitude and phase slope determination, we used only the most recent data, those since JD 2459641.5
up to JD 2460559.5, to avoid any possible small change due to the slight variation of the aspect angle in several years. From a linear fit of the reduced magnitudes versus phase angle, the $H_V$ determined is 5.40 $\pm$ 0.02 and the phase slope parameter is 0.18 $\pm$ 0.02 mag/deg. The fit was performed by removing outliers above and below 3$\sigma$. The $H_R$ and phase slope derived from the data are 4.85 $\pm$ 0.02 and 0.204 $\pm$ 0.014 mag/deg respectively. The $V-R$ is thus 0.55 $\pm$ 0.03.

The $H_V$ magnitude in the JPL Horizons system is 5.24, but this is assuming a fixed phase slope. Given the very heterogeneous sources of the data, this value is highly uncertain. Nevertheless, it is not too far from the value determined here.

In \cite{perna_photometry_2013}, a value of $H_V=5.36 \pm 0.13$ is reported in Table 3, based on their own observations, but this assumes a phase slope of 0.12 $\pm$ 0.06 mag/degree according to the table. However, when we correct their $M_V(1,1,1.1)$ values reported in their Table 2\footnote{Namely: 5.59 $\pm$ 0.04, 5.64 $\pm$ 0.03, 5.58 $\pm$ 0.04, 5.59 $\pm$ 0.04, 5.59 $\pm$ 0.04, 5.53 $\pm$ 0.05} and take the mean, the obtained value is 5.59 $\pm$ 0.02. Given that the observations were obtained at 1.1 degrees of phase angle, the final value is (using the reported phase slope) $H_V=5.46 \pm 0.07$, not 5.36. If we use our phase slope parameter of 0.18 mag/deg we obtain $H_V=5.40 \pm 0.03$, perfectly compatible with our own derived value. 

In \cite{alvarez-candal_absolute_2019} photometric data of \objname was also part of their data set. From an internal communication, they obtained values $H_V = 5.71 \pm 0.14$ and $H_R = 4.99 \pm 0.14$, which are also compatible with our results.

From our own photometric measurements, the peak-to-valley amplitude in the rotational light curve seems to be $\lesssim$ 0.1 magnitudes. Due to this small photometric variation, no rotational period could be confidently recovered from our data.

\section{Methodology}
\label{sec:method}

\subsection{The model}
\label{sec:model}

During the stellar occultation, \objname was at a geocentric distance of $\Delta = 35.532$ AU. Hence, the Fresnel scale $F = \sqrt{\lambda \Delta /2}$ at a wavelength of $\lambda = 600$ nm is $F = 1.3$ km. The stellar diameter can be estimated from the formulae in \cite{van_belle_predicting_1999}. Using the magnitudes in Table \ref{tab:star}, the angular diameter at the distance of \objname is 0.14 km assuming a main sequence star, 0.27 km assuming a variable star and 0.17 km assuming a super giant. The data from the occultation revealed a double star, so these values represent an upper limit for the individual diameters.

The shortest exposure time in our data set is 0.3 s, from the Artemis Observatory. In that time interval, the sky plane covered distance was 7.1 km, according to the mean shadow velocity of 23.6 km/s. This value is considerably larger than that of the Fresnel scale or any possible stellar diameter. As a result, our light curves are mainly dominated by exposure times. We consequently modelled the occultation with a purely geometric approach; considering the stars as point-like sources and with no diffraction effects.

\begin{figure}
\centering
\includegraphics[width=\hsize]{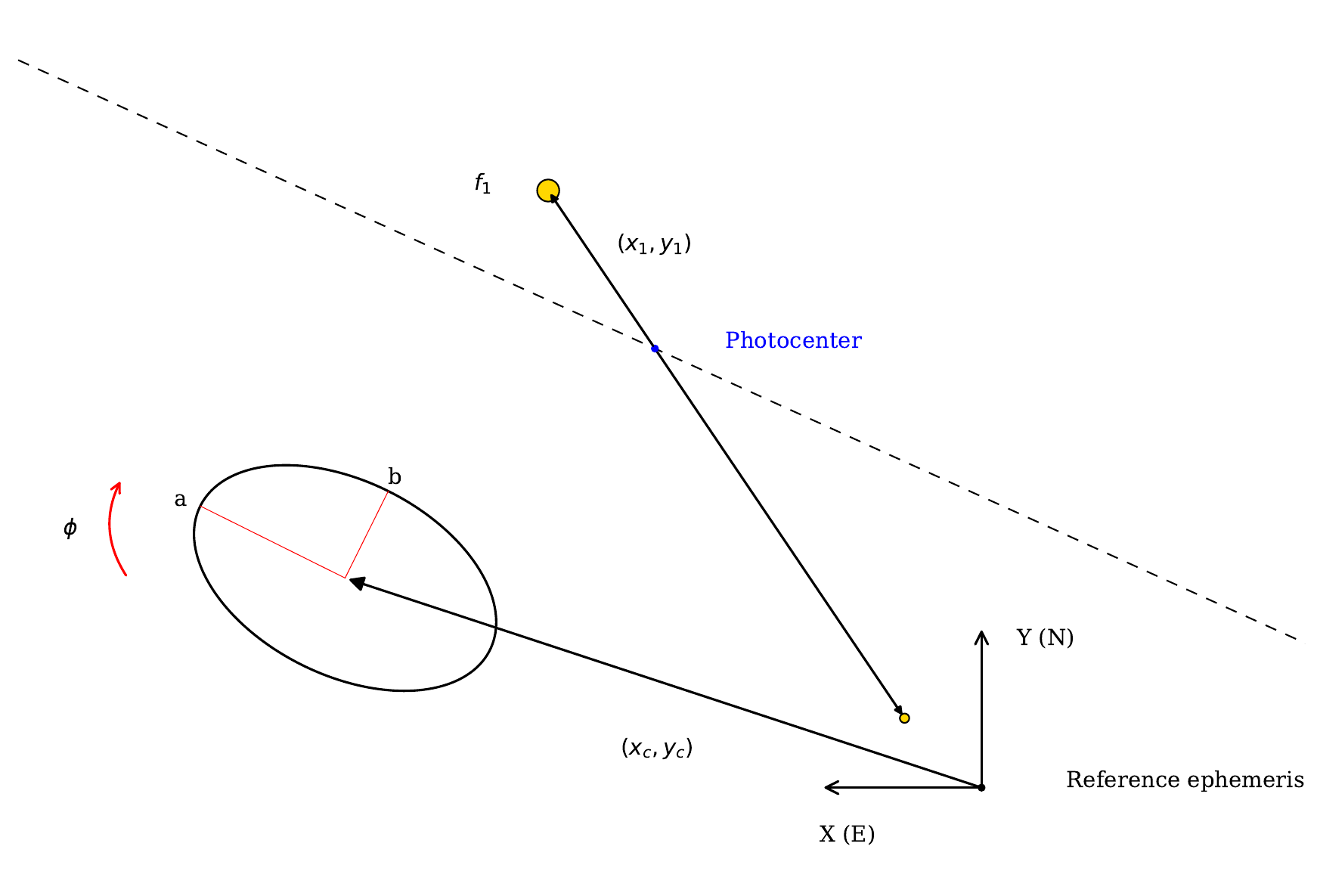}
\caption{Schematic representation of the model. The dashed line represents the path followed by the target star photocentre ($\alpha _{\text{ast}}, \delta _{\text{ast}}$ in Table \ref{tab:star}) on the sky plane, as seen from a certain site. The stars are represented by the yellow dots; the size of the dot is proportional to the brightness. The origin on the sky plane (black dot) is determined by the theoretical astrometric position of \objname according to our reference ephemeris. In this particular configuration of parameters, the site for which the chord is represented would observe the occultation, with a residual flux in the drop of $f_{\text{1}}$ (secondary star occulted, primary star visible).}
\label{fig:modelschema}
\end{figure}

The canvas for our modelling is the sky plane. This plane is perpendicular, at any instant, to the radius vector connecting the centre of \objname and the photocentre of the target stars (for a detailed explanation, see \citealt{elliot_radii_1978} and references therein). We set the origin of this plane to be the predicted barycentre of \objname according to some reference ephemeris. In our case, we choose as reference ephemeris the updated version from CAHA data. The Cartesian axes are oriented so that X corresponds to the eastwards direction and Y corresponds to the northwards direction (see Fig. 
 \ref{fig:modelschema}).

The projected shape of \objname will be modelled as an ellipse. In light of the observed data, we need to include the presence of two stars in the occultation (see discussion in Sect. \ref{sec:data_red}). Consequently, our model depends on the following 8 free parameters:
\begin{itemize}
    \item $a$ is the semi-major axis of the ellipse
    \item $b/a$ is the axis ratio ($b$ represents the semi-minor axis).
    \item $\phi$ is the tilt angle of the ellipse: from the East direction to the semi-major axis. This angle is restricted to the [-90$^{\circ}$ , 90$^{\circ}$] range, being positive when measured towards the North.

    \item ($x_c, y_c$) represents the position of the ellipse on the sky plane. These two parameters represent the offset of \objname with respect to the reference ephemeris. Given that the offset is approximately constant during the short duration of the occultation, its value provides a new astrometric measurement of the \objname for further refining its orbit.
    \item $f_1$ represents the normalised flux of the brightest star (primary star). As discussed in Sect. \ref{sec:data_red}, the expected flux drop was 99.5 \% considering a single star. Since our light curve with the highest S/N is the one from Artemis with S/N=30, this residual brightness from \objname will not be distinguishable from noise and is neglected. Therefore, the normalisation of the light curves implies $1 = f_1 + f_2$, where $f_2 = 1 - f_1$ is the normalised flux of the fainter star (secondary star).

    \item ($x_{1}, y_1$) define the position of the primary star on the sky plane, taking the photocentre as origin: $(x_{1}, y_1) =\va{r}_{\text{star 1}} - \va{r}_{\text{photocentre}}$. The position relative to the photocentre for the secondary star $(x_2, y_2)$ is now completely determined by
    \begin{equation}
        (x_2, y_2) = -\frac{f_1}{1 - f_1}(x_1, y_1).
        \label{eq:starPositions}
    \end{equation}
\end{itemize}

\subsection{The Bayesian approach}
\label{sec:stat_approach}

The traditional analysis of a stellar occultation is usually performed as follows. Firstly, the obtained light curves are identified as positive (flux drop detected) or negative (no detection). In most cases, diffraction or extensive source effects are negligible (see Sect. \ref{sec:model}), so a square-well is fitted to all the positive light curves. This determines the ingress and egress times (beginning and end of the drop). Then, they are compared to synthetic times produced by certain shape models. The differences between synthetic and measured times are minimised following an iterative $\chi ^2$ approach. Many TNOs have been successfully modelled with this procedure \citep{santos-sanz_physical_2022, pereira_two_2023}. 

However, this methodology can be impractical for moderate to low signal-to-noise ratios in some of the light curves, as is the case in the present work. This situation can be aggravated in the scenario of a double star being occulted. If one of the stars is significantly fainter, stations detecting only the occultation of this star might not be able to distinguish the small flux drop from the noise. Therefore, we adopted an approach where each data point from each light curve is treated individually, comparing them to synthetic light curves produced by a certain model. In this way, a flux drop smaller than noise level due to the occultation of the faint star is naturally covered, without the need to decide if the occultation was actually detected or not. This is the case of TAR2 data. Ingress and egress times should have been computed in a classic approach. Clearly, this is not an easy task when the data quality is not optimal (see Fig. \ref{fig:lcs}). Worse still, in the absence of Artemis data, the TAR2 light curve could have been mistakenly identified as negative, yielding to misleading conclusions. In this aspect, our methodological approach is advantageous, as no a priori assumptions need to be made about the nature of the light curves. 

Although this is not the present case, the situation in which a single station observes the blended occultation of both stars would be treated the same way with this approach. In contrast, determining ingress and egress times for each star would imply fitting increasingly complex stepped functions, which might be cumbersome in noisy data sets.

To fit the synthetic data curves produced by our model to the observed ones, we follow a methodology similar to that of \cite{leiva_stellar_2020} or \cite{strauss_sizes_2021}. We use Bayesian statistics to determine a posterior probability density function (posterior pdf or simply posterior) for the parameters described before under our adopted model. The aim is to characterise this posterior. This approach is advantageous because it provides a better understanding of parameter correlations, which are likely to appear in double star occultations. By characterising the full posterior distributions, we are also able to comprehensively quantify uncertainties \citep{gregory_bayesian_2005}. 

The posterior pdf \posterior is constructed from the likelihood function \lhood, a prior distribution of the model parameters \prior and a normalisation factor $P(D)$:
\begin{equation}
    P(\theta | D) = \frac{\mathcal{L}(D | \theta) P(\theta)}{P(D)}.
\end{equation}

Here $D$ represents the obtained data set, comprising each normalised flux value $f_i$ for each light curve. $\theta$ is our 8-dimensional parameter vector. The likelihood measures the plausibility of observing the data set $D$ given the parameter values $\theta$. The prior distribution reflects our a priori knowledge about the model parameters. In this work, we consider the simplest case where \prior is the product of the marginal prior distributions for each parameter (no correlations).

Modelling the normalised flux errors $\sigma _i$ as normally distributed and uncorrelated, the likelihood function is given by \citep{gregory_bayesian_2005}

\begin{equation}
\label{eq:lhood}
    \mathcal{L}(D|\theta) = (2\pi)^{-N/2} \qty{\displaystyle \prod _{i=1} ^N \frac{1}{\sigma _i}} \times \qty{\exp \qty[-\displaystyle \sum _{i=1} ^N \frac{(f_i - m(t_i|\theta))^2}{2\sigma _i ^2}]},
\end{equation}

where $N$ is the total number of data points and $m(t_i|\theta)$ is the modelled light curve value under the parameters $\theta $ at time $t_i$, corresponding to the data point $f_i$.

To characterise the posterior pdf, we applied a Markov chain Monte Carlo (MCMC) method to obtain a representative sample from \posterior. We adopt the use of Python package \texttt{emcee} 3.1.6 \citep{foreman-mackey_emcee_2013}, which implements the affine invariant sampling algorithm described in \cite{goodman_ensemble_2010}.

In our samplings, we used $n_w = 512$ random walkers, which are initially distributed according to the selected prior. The sampler is initially run for $n_{\text{burn}}$ steps to ensure convergence of the sampling. This is the \emph{burn-in} phase, where the walkers explore the parameter space. We checked that the number of steps in this phase is at least 10 times larger than the estimated auto-correlation time in each parameter, to ensure that we are not sampling far from convergence. From the last position of the burn-in phase, the sampler is run for $n_{\text{iter}}$ additional steps to obtain a representative sample. The final sample size is $n_w \times n_{\text{iter}}$ (we discard the burn-in phase). 

Our MCMC samplings were performed with $n_{\text{burn}}=3000$ and $n_{\text{iter}}=1000$. Since \posterior does not show signs of being multi-modal (just one set of parameters satisfactorily explains the data), it is sensible to present the median of the corresponding marginal distributions as the nominal values of each parameter. The provided errorbars are taken from 68 \% maximum-density credible intervals. That is, the shortest interval containing 68 \% of the sample elements, corresponding to the 1$\sigma$ region.

\section{Occultation results}
\label{sec:results}

\subsection{Bayesian analysis: Prior selection and results}
\label{sec:MCMC_results}

In Bayesian analysis, prior selection is a sensitive matter as it can significantly influence the posterior in cases of low S/N data. The only a priori information we have about \objname is the published result from thermal data. Population-wise, we also have certain information from published TNO occultations about the shape distribution. We report the results of two different samplings taking these into account. We aim to assess the relative influence of the priors and occultation data in the posterior.

Our preferred sampling (sampling 1) will have uniform priors in all parameters. This is the most uninformative choice and serves only a practical purpose of limiting the parameter-space region the walkers must explore. We selected this as our preferred sampling, since we intend to compare the results on the object size from thermal data reported in \cite{santos-sanz_tnos_2012}. Including these as priors would bias our sampling beforehand, invalidating the comparison.

Our sampling 2 includes as priors the area-equivalent diameter from thermal data and shape distribution of TNOs from stellar occultations. Since $D_{\text{eq}}$ is not directly a parameter of our model, we need to introduce prior distributions in $a$ and $b/a$ so that the resulting prior is compatible with a normal centred in the nominal $D_{\text{eq}}$ value and $\sigma$ equal to the reported uncertainty. To do so, we considered that the best practical choice for the $b/a$ prior would be a normal fit to the observed $b/a$ values in other TNO occultations. Then, the prior in semi-major axis can be chosen accordingly.

In Table \ref{tab:boas} we collect all the published values for projected axis ratio in TNO stellar occultations. We acknowledge that the sample of TNOs with published occultations is biased towards the largest objects. This highlights the need to accumulate more data if we intend to perform statistically significant analyses. 

\begin{table*}
\caption{All published TNO projected shape axis ratios from occultations.}
\label{tab:boas}
\renewcommand{\arraystretch}{1.3} 
\centering
\begin{tabular}{lrrrp{2.1 cm}}
\hline \hline
Name & Class & $b/a$ & $D_{\text{eq}}$ (km) & References\\
\midrule
2003 UY117 & Resonant 5:2 & 0.65 $\pm$ 0.15 & 230 $\pm$ 26 & 1\\
\textbf{2007 OC10} & \textbf{Detached} & \textbf{0.58 $\pm$ 0.16} & {$\mathbf{330} ^{\mathbf{+56}}_{\mathbf{-55}}$} & \textbf{This work}\\
Huya & Plutino & 0.89 $\pm$ 0.03 & 411 $\pm$ 10 & 2 \\
2002 TC302 & Resonant 5:2 & 0.85 $\pm$ 0.04 & 500 $\pm$ 19 & 3\\
2003 VS2 \tablefootmark{a}  & Plutino & 0.79 $\pm$ 0.02 & 519 $\pm$ 10 & 4\\
2003 VS2 \tablefootmark{a}  & Plutino & 0.81 $\pm $ 0.08  & 570 $\pm $ 31 & 5\\
G!kún\textdoublepipe\textprimstress hòmdímà & Detached & 0.89 $\pm $ 0.08  & 641 $\pm $ 28 & 6\\
2003 AZ84 \tablefootmark{a}  & Plutino & 0.66$\pm$ 0.11 & 692 $\pm$ 59 & 7\\
Varda & Resonant 15:8 & 0.93 $\pm$ 0.05 & 740 $\pm$ 20 & 8\\
2003 AZ84 \tablefootmark{a} & Plutino & 0.946 $\pm$ 0.005 & 764 $\pm$ 6 & 7\\
2002 MS4 & Hot classical & 0.93$\pm$ 0.05 & 794 $\pm$ 25 & 9\\
Quaoar & Hot classical & 0.880 $\pm$ 0.013 & 1087 $\pm$ 11 & 10\\
Haumea & Resonant 12:7 & 0.668 $\pm$ 0.015 & 1393 $\pm$ 16 & 11\\
Makemake & Hot classical & 0.95 $\pm$ 0.03 & 1466 $\pm$ 49 & 12 \\

\bottomrule
\end{tabular}
\tablefoot{The area-equivalent diameter for each fitted ellipse is also included. We excluded cases where only a circle was fitted. We did not include the remarkable case of Arrokoth \citep{buie_size_2020} due to its binary nature. Dynamical classification according to \cite{volk_dynamical_2024}. \tablefoottext{a}{\footnotesize These objects have two different occultations published with fitted ellipses. We treat them as separate data points although they are clearly not independent.}}
\tablebib{
(1) \cite{kretlow_physical_2024};
(2) \cite{santos-sanz_physical_2022};
(3) \cite{ortiz_large_2020};
(4) \cite{vara-lubiano_multichord_2022};
(5) \cite{benedetti-rossi_trans-neptunian_2019};
(6) \cite{benedetti-rossi_results_2016};
(7) \cite{dias-oliveira_study_2017};
(8) \cite{souami_multi-chord_2020};
(9) \cite{rommel_large_2023};
(10) \cite{pereira_two_2023};
(11) \cite{ortiz_size_2017};
(12) \cite{ortiz_albedo_2012}.
}
\end{table*}

The fitted truncated normal between 0 and 1 for these axis ratio values is $N(\mu = 0.889, \sigma = 0.143)$ (see Appendix \ref{ap:priors} for details). If a normal prior $N(\mu = 169\ \text{km}, \sigma = 17\ \text{km})$ is chosen for $a$, the resulting a priori distribution for $D_{\text{eq}}$ resembles the desired distribution from thermal data $N(\mu = 309\ \text{km}, \sigma = 37\ \text{km})$: The mean is $\mu = 308.1$ km, the median is $M = 308.2$ km and the standard deviation is $\sigma = 37.1$ km. To further confirm the compatibility, we performed a Kolmogorov-Smirnov test comparing a sample of this a priori $D_{\text{eq}}$ distribution against $N(\mu = 309\ \text{km}, \sigma = 37\ \text{km})$. With a sample size of 1000, the test yields a p-value=0.61. Therefore, it is unjustified to reject the null hypothesis: The sample comes from the desired distribution.

\begin{figure*}
\centering
\includegraphics[width=0.8\hsize]{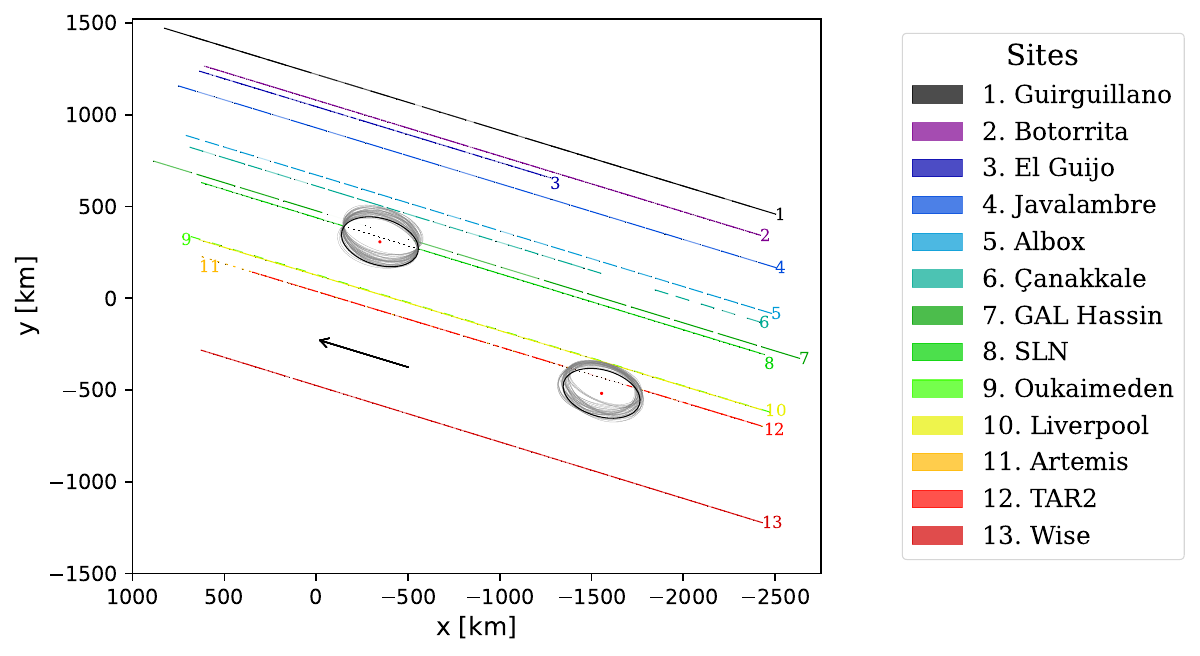}
\caption{Occultation data plotted on the sky plane with the results of sampling 1. The X- and Y-axes are pointing towards east and north, respectively. The arrow indicates the relative direction of motion of the occulted stars. Each line segment represents an exposure. The transparency of each segment corresponds to the normalised flux measured in that integration (see Fig. \ref{fig:lcs}). No transparency means flux = 1. In the light curve from Çanakkale there is an abnormally long dead-time between two concrete exposures. The corresponding blank space should not be mistaken for a flux drop. The pair of black ellipses represent the nominal solution reported in Table \ref{tab:mainresults} for sampling 1. Their centres are marked with red dots. The thinner grey ellipses represent the 1$\sigma$ uncertainties from the extracted sample, as discussed in the text.  }
\label{fig:chords&result}
\end{figure*}

For sampling 1 and sampling 2, we selected uniform priors in the position parameters $(x_c, y_c)$. We know that the actual position ($x_c, y_c$) of \objname on the sky plane has to be between the two flux-drop areas. And it should be closer to the drop from SLN and GAL Hassin, sites where the occultation by the primary star was detected. Since $f_1 \simeq 0.75$, from Eq. \ref{eq:starPositions} we can infer that ($x_c, y_c$) should be three times closer to this flux drop area than the one from Artemis and TAR2 (secondary star). With this in consideration and visually inspecting the light curves plotted on the sky plane (see Fig. \ref{fig:chords&result}), we choose an uniform prior in the region $[-900, -300]\ \text{km} \times [-500, 400]\ \text{km}$, confidently covering the actual value of ($x_c, y_c$). This limits the futile exploration of regions where the ellipse positions on the sky plane are far away from the actual flux drop areas.

The situation for ($x_1, y_1$) is similar to that of ($x_c, y_c$), since the double nature of the star was unknown prior to the occultation. Hence, we base the prior selection in practical purposes. From visual inspection, the primary star position relative to the photocentre should be in the $x_1, y_1 < 0$ region, to cause an occultation visible from SLN and GAL Hassin (see Fig. \ref{fig:modelschema}). Consequently, a conservative uniform prior is chosen in the region $[-150,-500]\ \text{km} \times [0,-350]\ \text{km}$ to limit the walkers from exploring unnecessary parameter space volume.

 For the rest of parameters that have not been discussed, the priors are the same for sampling 1 and sampling 2:

\begin{itemize}
 \item The possible orientation of the projected shape was unknown a priori, hence we set a completely uninformative prior for $\phi$: U$[-90^\circ, 90^\circ]$.

\item As the star was an unknown double, there is no a priori information about the possible value of $f_1$. Hence, we set an uninformative prior U[0.5, 1].
\end{itemize}

\begin{table*}
\caption{Results from our MCMC samplings.}
\label{tab:mainresults}
\centering
\renewcommand{\arraystretch}{1.3} 
\begin{tabular}{c c c c c} 

\hline\hline
Parameter & \multicolumn{2}{c}{\textbf{Sampling 1}} & \multicolumn{2}{c}{Sampling 2} \\ 
\cmidrule(lr){2-3} \cmidrule(lr){4-5}
 & \textbf{Prior} & \textbf{Result} & Prior & Result \\ 
\hline
$x_c$ (km)      & \textbf{U[-900, -300]} & \textbf{-653$^{+66} _{-33}$}   &  U[-900, -300] & -652$^{+40} _{-31}$  \\
$y_c$ (km)      & \textbf{U[-500, 400]}   & \textbf{98$^{+58} _{-180}$}    & U[-500, 400] & 106$^{+170} _{-100}$  \\
$a$ (km)        & \textbf{U[100, 300]}    & \textbf{215$^{+10} _{-7}$}     & N(169, 17) & 211$^{+11} _{-5}$  \\
$b/a$           & \textbf{U[0.0, 1.0]}    & \textbf{0.58$^{+0.16} _{-0.16}$} & N(0.889, 0.143; 0,1) & 0.68$^{+0.11} _{-0.11}$  \\
$\phi$ (deg)    & \textbf{U[-90, 90]}     & \textbf{16$^{+12} _{-11}$}     & U[-90, 90] & 17$^{+9} _{-8}$  \\
$f_1$           & \textbf{U[0.5, 1.0]}    & \textbf{0.748$^{+0.011} _{-0.012}$} & U[0.5, 1.0] & 0.747$^{+0.012} _{-0.014}$  \\
$x_{\text{1}}$ (km) & \textbf{U[-500, -150]} & \textbf{-305$^{+31} _{-38}$}   & U[-500, -150] & -307$^{+24} _{-21}$  \\
$y_{\text{1}}$ (km) & \textbf{U[-350, 0]} & \textbf{-209$^{+22} _{-146}$}   &  U[-350, 0] & -206$^{+72} _{-27}$  \\
\hline
\end{tabular}
\tablefoot{Sampling 1 is our preferred sampling, with all uniform priors. The second sampling has priors that take into account the projected axis ratio distribution for published occultations in addition to the reported area-equivalent diameter in \cite{santos-sanz_tnos_2012} (see Fig. \ref{fig:cornerplot} for an overview of the posterior pdfs from which these values were obtained; see text).}
\end{table*}

We represent the resulting samples in a pairwise marginal plot in Fig. \ref{fig:cornerplot}. From this representation it is clear that the samplings show no signs of multi-modality, so there is no degeneracy in the solutions. The clearest correlations appear between the positional parameters $x_c$, $x_1$ and $f_1$, which is not unexpected. Slight variations of the position of \objname and the primary star can produce the same shadow, as long as $f_1$ is adjusted so that the secondary star shadow also stays in place. However, since the value $f_1$ is accurately determined by Artemis light curve, this variation is very well constrained. Another (less marked) correlation can be found between $b/a$ and $y_c$: If the position of \objname on the sky plane was further south, a moderately rounder projected shape would be admissible.

The nominal parameter values with the corresponding uncertainties for both samplings are given in Table \ref{tab:mainresults}. Figure \ref{fig:chords&result} offers a qualitative visualisation of the result of our preferred sampling 1. A sub-sample of 100 parameter vectors is extracted randomly from the full posterior sample. Those belonging to the 68 \% credible region of each parameter space are shown by drawing the flux drop area (i.e. the projected elliptical limb) on the sky plane expected for this parameter configuration. The expected shadows for the nominal parameters in Table \ref{tab:mainresults} are also represented for reference.

Along both samplings, the obtained values for the model parameters are compatible within the provided (1$\sigma$) credible intervals. This is an indication that the sampling is heavily influenced by the obtained occultation data, and not by the chosen priors. 

The axis ratio parameter is the one suffering from greater variation between sampling 1 and sampling 2. This is expected since the prior in $b/a$ for sampling 2 gives preference to rounder objects, favouring a higher median in the posterior marginalised distribution for $b/a$. However, this axis ratio prior is based on a scarce sample, without significant representation of objects of the same size or dynamical class as \objname. Therefore, we prefer to report sampling 1 result as more realistic and unbiased.

\begin{figure}
\centering
\includegraphics[width=\hsize]{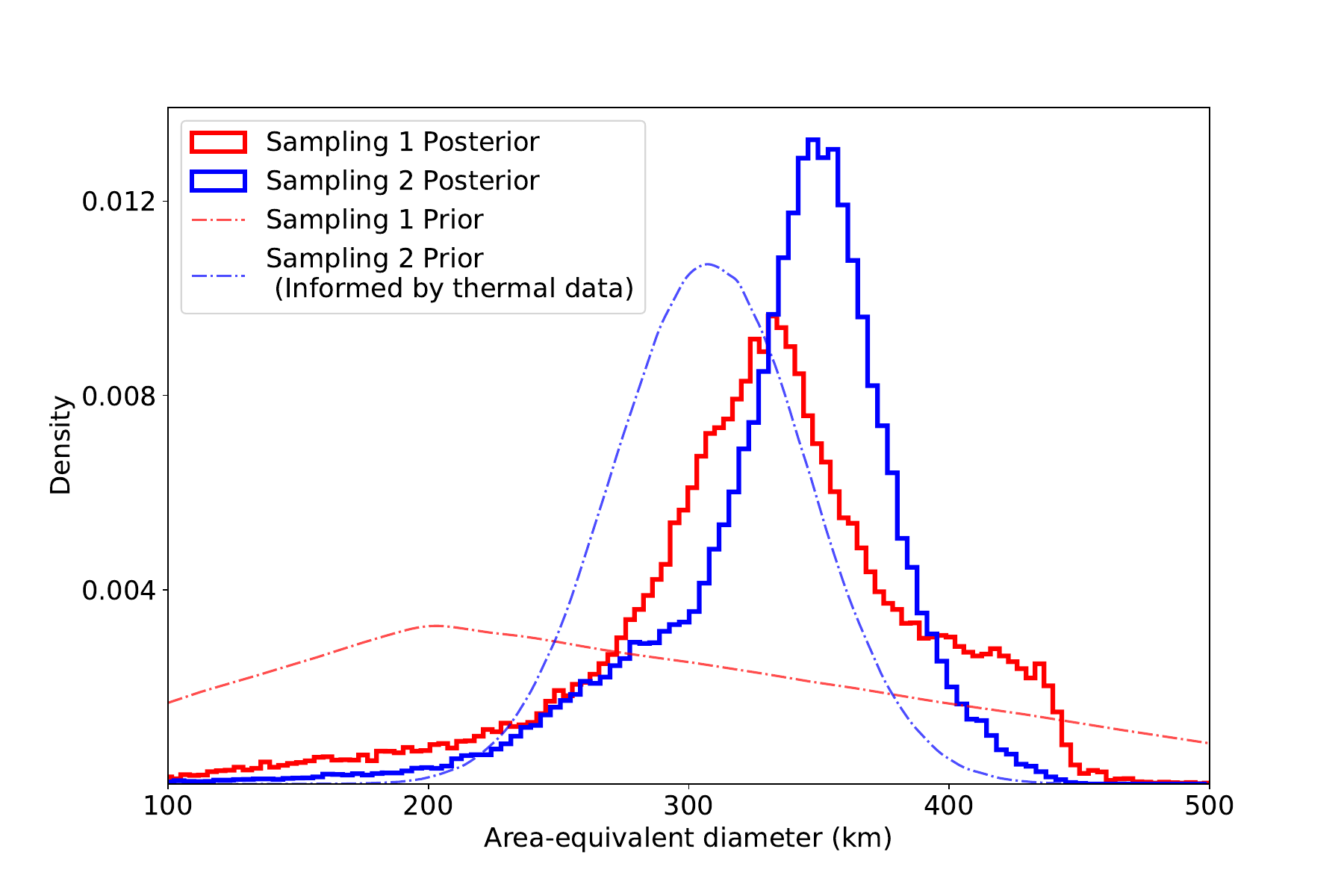}
\caption{Histogram comparison of the posterior samples in $D_{\text{eq}}$. The priors are plotted for reference (dashed lines). Sampling 1 is our preferred sampling, with uniform priors in all our model parameters. However, as $D_{\text{eq}}$ is not directly a model parameter, the prior in this value is not uniform. In sampling 2, the prior in $D_{\text{eq}}$ is equivalent to the value reported in \cite{santos-sanz_tnos_2012} from thermal data. This prior pdf peaks at a higher $D_{\text{eq}}$ value than the prior pdf of sampling 1. Consequently, the posterior pdf in sampling 2 also peaks at a higher value than in sampling 1. However, both distributions yield compatible results within 1$\sigma$ (68\% credible intervals). Moreover, the result from sampling 1 is also compatible with the result from thermal data (sampling 2 prior). }
\label{fig:DeqsComparison}
\end{figure}

The differences in axis ratio due to different priors are translated in a difference of the obtained area-equivalent diameters: $D_{\text{eq}}=330 ^{+56}_{-55}$ km in sampling 1 and $D_{\text{eq}}=343 ^{+40}_{-34}$ km in sampling 2. In Fig. \ref{fig:DeqsComparison}, it is shown how the different prior distributions in $D_{\text{eq}}$ affect the resulting sample. The combination of $a\sim U[100, 300]$ km and $b/a \sim U[0.0, 1.0]$ (sampling 1) results in a $D_{\text{eq}}$ prior with a peak at $\sim$ 200 km, but very long tails compared to the $N(309,37)$ km prior distribution in sampling 2. Despite the notable differences in these priors, the results obtained are compatible within the credible interval $1\sigma$. Again, this supports our claim that the posterior density is primarily influenced by the likelihood (determined by the occultation data) and not the priors.

\subsection{Projected axis ratio of the object}

\label{sec:ShapeResult}

The determination of the projected axis ratio of the object $b/a = 0.58 \pm 0.16$ (or equivalently, $b=141^{+24}_{-23}$ km) is of great relevance. This adds another data point to the scarce data set on projected TNO shapes (see Table \ref{tab:boas}). The importance of this value is twofold.

Firstly, the scarcity of the data is further emphasised when dividing the sample according to dynamical classes. This division might be of interest if we seek to identify differences in the shape distribution according to origin, since we know that the evolutional history varies between dynamical classes \citep{nesvorny_dynamical_2018}. As stated previously, this is only the second detached TNO to be measured with stellar occultations.

Secondly, \objname is only the third-smallest TNO with a measured projected shape through occultations. This is of remarkable interest, as shapes are expected to depend notably on size \citep{tancredi_which_2008}. Theoretically, there should be a certain critical diameter above which shapes should be governed by hydrostatic equilibrium \citep{chandrasekhar_ellipsoidal_1987}. However, there is no empirical evidence of where exactly this limit is located. In fact, even the existence of a critical diameter is to be confirmed, since some large ($D_{\text{eq}} > 500$ km) TNOs depart from hydrostatic shapes of equilibrium \citep{ortiz_size_2017, vara-lubiano_multichord_2022}. Therefore, it is important to explore the lower end of the size parameter space in order to acquire sufficient data to test this hypothesis with some statistical significance.

\subsection{Surface properties: Geometric albedo and beaming parameter}
\label{sec:beaming}

In this section we combined our occultation results from the elliptical model with photometric and thermal measurements to refine our knowledge about the geometric albedo $p_V$ and beaming parameter $\eta$. We contrast them with the values derived in the literature from the TNOs are Cool survey \citep{santos-sanz_tnos_2012, lellouch_tnos_2013}.

We acknowledge that the projected area (and consequently $D_\text{eq}$) at the instant of the occultation could differ from the one at the epoch of thermal measurements. Due to the absence of information on the three-dimensional shape of \objname and, in particular, its rotation axis, we can only statistically estimate the variation in projected area caused by rotation and changes in aspect angle. The statistical analysis is performed in Appendix \ref{ap:area_change}. While the difference in projected area between the two epochs may be significant, for the purposes of this study, we assume that the projection observed during the occultation and thermal measurements remains the same.

The geometric albedo and beaming parameter values obtained in \cite{santos-sanz_tnos_2012, lellouch_tnos_2013} from thermal data come from a modified NEATM model \citep{harris_thermal_1998}. The only difference with the original NEATM model is that zero phase angle is assumed. The NEATM model assumes a spherical body in instantaneous equilibrium with the solar radiation, in the limit of zero thermal inertia. A so-called beaming parameter $\eta$ is introduced to account for surface roughness and the fact that the thermal inertia is non-zero. The two unknown parameters that completely determine the thermal emission in the NEATM model are the geometric albedo $p_V$ and the beaming parameter $\eta$. The absolute magnitude $H_V$ of the object is also needed to determine the diameter of the sphere when combined with $p_V$. The absolute magnitude is obtained from independent photometric measurements.

Within the TNOs are Cool survey, the thermal flux from \objnamelong was observed with \textit{Herschel}/PACS at three different wavelengths: 70 {\textmu}m, 100 {\textmu}m and 160 {\textmu}m. As discussed in \cite{santos-sanz_tnos_2012}, only these three data points were considered insufficient to fit $p_V$ and $\eta$ simultaneously. Hence, the authors adopted an approach in which the beaming parameter is fixed to $\eta = 1.20 \pm 0.35$ (the mean value known at the time for TNOs; \citealt{stansberry_physical_2008}), so they fit $p_V$ as the only free parameter. 

Using the absolute magnitude derived from our own observations of $H_V = 5.40 \pm 0.02$ (see Sect. \ref{sec:photometry}), the obtained value for the geometric albedo is $p_V = 11.2 ^{+2.1}_{-5.0}$ \%. This is consistent with $p_V = 12.7^{+0.4}_{-2.8}$ \% from \cite{santos-sanz_tnos_2012}. A $H_V$ value of $5.43 \pm 0.10$ was used on their work, similar to that derived in the present paper. If we adopt their value of absolute magnitude, the variation in our result is negligible (see Table \ref{tab:thermal_comparison}). 

\cite{lellouch_tnos_2013} reanalysed the same thermal data for \objname reported in \cite{santos-sanz_tnos_2012} but fitting both parameters $p_V$ and $\eta$, instead of using a fixed $\eta$ value (see Table~\ref{tab:thermal_comparison}). The uncertainties are considerably larger than in \cite{santos-sanz_tnos_2012}, and $\eta$ is barely constrained. The beaming parameter for \objname is effectively unknown. Occultation data can help constrain this value. Because we have now independently determined $p_V$, we can fit $\eta$ as the only free parameter. For comparability, we adopt the same modified NEATM model, phase integral approximation, and fitting method used in \cite{santos-sanz_tnos_2012} and \cite{lellouch_tnos_2013}.

Adopting the most recent and precise absolute magnitude value $H_V = 5.40 \pm 0.02$ derived in this work, the resulting beaming parameter is $\eta = 1.42^{+0.75}_{-0.58}$. The value obtained for $H_V = 5.43 \pm 0.10$ is also reported in
Table \ref{tab:thermal_comparison} for completeness. There is a considerable reduction in the uncertainty of $\eta$ with respect to the value reported in \cite{lellouch_tnos_2013}. Our result shows that \objname has a beaming parameter compatible with the mean of the population. However, the relative error is still $\sim$ 50 \% in $\eta$, preventing us from drawing any further conclusions.

\begin{table}
\caption{Geometric albedos, beaming parameters, and area-equivalent diameters for \objname from our work and thermal data of the TNOs are Cool survey. }
\label{tab:thermal_comparison}
\renewcommand{\arraystretch}{1.4}
\setlength{\tabcolsep}{3.5pt}
\centering
\begin{tabular}{p{1.6 cm}p{1.1 cm}p{1.7 cm}p{1.3 cm}p{2.0 cm}}
\hline \hline
 $H_V$ \tablefootmark{a} & $p_V$ (\%) & $\eta$ & $D_{\text{eq}}$ (km) & References \\
\midrule
 \textbf{5.40 $\pm$ 0.02} & \textbf{11.2$^{+2.1}_{-5.0}$} \tablefootmark{b} & \textbf{1.42$^{+0.75}_{-0.58}$} \tablefootmark{c} & \textbf{330 $^{+56}_{-55}$} & \bf{This work} \\
  $5.43 \pm 0.10$ & $10.9 ^{+2.5}_{-5.0}$ \tablefootmark{b} & $1.39 ^{+0.68}_{-0.56}$ \tablefootmark{c}& $330 ^{+56}_{-55}$ & This work ($H_V$ from 1)\\
 $5.43  \pm 0.10$ & $12.7^{+4.0}_{-2.8}$ & $1.20 \pm 0.35$ \tablefootmark{d} & $309 \pm 37$ & 1 \\
 $5.43  \pm 0.10$ & $13.0^{+9.5}_{-5.6}$ & $1.17^{+1.11}_{-0.64}$ & $306 ^{+93}_{-72}$ & 2 \\
 
\bottomrule

\end{tabular}
\tablefoot{All values considering a single object.
\tablefoottext{a}{\footnotesize From independent photometric measurements.}
\tablefoottext{b}{Obtained from combining the corresponding $H_V$ value with the occultation-derived $D_{\text{eq}}$.}
\tablefoottext{c}{Fitted in the NEATM with $p_V$ fixed from the occultation.}
\tablefoottext{d}{\footnotesize Not fitted, value fixed to the mean $\eta$ of the TNO population.}
}
\tablebib{
(1) \cite{santos-sanz_tnos_2012};
(2) \cite{lellouch_tnos_2013}.
}
\end{table}

\subsection{Double star characterisation}
\label{sec:star_charact}

As a by-product of the shape analysis of \objname, we have constrained the relative position of the occulted stars at the epoch of the occultation: 2022-08-22T01:32:37 UTC. The fitting of ($x_1,y_1$) and $f_1$ completely determines the relative position of both stars. From our preferred sampling, we computed an angular separation of $57^{+4}_{-11}$ mas and a position angle of the secondary star with respect to the primary star (measured from north towards the East) of $\theta = 56 ^{+3}_{-17}$ degrees.

In addition, the fitting of $f_1$ allows us to conclude that the instantaneous magnitude difference between stars was $1.18 ^{+0.07} _{-0.07}$ magnitudes.

\subsection{Astrometric constraints}
\label{sec:astrometric_constr}

Occultations are a source of mas-level astrometry for TNOs \citep{rommel_stellar_2020}. In this particular case, the occultation also enables us to constrain the astrometric position of both stars. All these astrometric constraints rely on the precise determination of the double star photocentre at the occultation epoch (see Table \ref{tab:star}). To compute this position ($\alpha _{\text{ast}}$, $\delta _{\text{ast}}$) the parallax and proper motion needs to be taken into account. However, the duplicity nature of the star was unknown when computing these values, which might be a source of systematic error. As discussed in \cite{leiva_stellar_2020}, in case both stars are gravitationally bound, then the \textit{Gaia} DR3 proper motion reflects a combination of the star system barycentric motion and the movement of the photocentre relative to the barycentre. On the contrary, if the stars are just aligned by chance, then \textit{Gaia} DR3 proper motion represents a combined effect of the individual proper motion and parallax of each star. Trying to discern between both cases is neither possible with the occultation data nor in the scope of this work.

We report the instantaneous astrometric positions of \objname and both stars in Table \ref{tab:astrometry}. We note that our model assumes a fixed position for the photocentre with no uncertainty. Therefore, the uncertainties in Table \ref{tab:astrometry} combine the uncertainties from the occultation sampling result and those from the \textit{Gaia} astrometric catalogue ($\alpha _{\text{ast}}$, $\delta _{\text{ast}}$ in Table \ref{tab:star}). The latter ($\mathcal{O}(0.1\ \text{mas})$) are assumed to be normally distributed and independent, which is a fair approximation since the dominating source of error comes from the sampling ($\mathcal{O}(1-10\text{mas})$).

\begin{table*}[h]
\caption{Astrometric constraints derived from the occultation analysis.} 
\label{tab:astrometry}
\centering
\begin{tabular}{p{6.6 cm}p{6.5 cm}@{}}
\toprule \toprule

Epoch of the reported astrometry & 2022-08-22T01:32:37 UTC \\ [0.1cm]
\midrule
Primary star & Secondary star \\ [0.1cm]
\midrule
$\alpha _1$= 22:08:33.0672 $\left( ^{+1.4}_{-1.6}\  \text{mas} \right)$ & $\alpha _2$= 22:08:33.0703 $\left( ^{+3.2}_{-5.4}\  \text{mas} \right)$\\ [0.1cm]
$\delta _1$= +12:05:55.6714 $\left( ^{+1.4}_{-5.5}\  \text{mas} \right)$ & $\delta _2$= +12:05:55.7035 $\left( ^{+1.6}_{-19.4}\  \text{mas} \right)$ \\ [0.2cm]

\midrule
\multicolumn{2}{c}{Object astrometric position (geocentric)} \\ [0.1cm]
\midrule
$\alpha _{\text{\objname}}$ & 22:08:33.0652 $\left( ^{+2.4}_{-1.9}\  \text{mas} \right)$ \\[0.1cm]
$\delta _{\text{\objname}}$ & +12:05:55.7402 $\left( ^{+2.7}_{-7.0} \ \text{mas}\right)$ \\ [0.1 cm]
\bottomrule
\end{tabular}
\tablefoot{ Considerations for the occulted stars and \objname astrometry are discussed in Sect. \ref{sec:astrometric_constr}.}
\end{table*}

\section{Comparison with thermal data. Upper limit on the size of a putative satellite }
\label{sec:satellite}

There are no secondary flux drops in the occultation light curves that suggest a secondary object, but Fig. \ref{fig:chords&result} shows that the sky plane area covered by the data has about 1500 km in width, with gaps between sites as large as 500 km. 
A secondary object in the gaps between observing sites or outside the band covered by the occultation data would remain undetected.
The thermal data, taken with \textit{Herschel}/PACS, have an angular resolution of 5 arcseconds \citep{Poglitsch2010}. The angular resolution corresponds to about 130000 km at the object's distance during the thermal measurements \citep{santos-sanz_tnos_2012}, a resolution too coarse to detect any of the known trans-Neptunian binaries \citep{Noll2020}. 
There are no published imaging data from the Hubble Space Telescope (HST), which could reach resolutions of about 1000 km at the distance of the Kuiper belt. 
Considering the limited angular resolution of the occultation and thermal imaging data, we cannot directly rule out the presence of a yet undetected secondary object. 

Fitting a single-object model to a hypothetical double-object to the thermal measurements will result in a systematically larger thermal diameter to account for the additional flux from the secondary. Neglecting differences due to changing geometry (see discussion in Sect. \ref{sec:beaming} and Appendix \ref{ap:area_change}), an unresolved secondary object will show up as a larger thermal diameter with respect to the occultation data. This has been previously proposed for the TNOs 2002~MS4 and 2002~TC302 \citep{rommel_large_2023, ortiz_large_2020}. Table \ref{tab:thermal_comparison} shows that the area-equivalent diameter from our elliptical model is compatible, within error bars, with the thermally derived diameter values. The relatively large uncertainties in the diameter from both independent measurements do not discard cases of a larger thermal diameter than the occultation diameter. Thus, we cannot categorically rule out a secondary, although the similarity in both thermal and occultation values allows us to constrain its maximum size. 

In this section we adopt a novel statistical approach to derive the upper limit on the size of a hypothetical secondary object, using the occultation-derived diameter for the primary and thermal data to constrain the diameter of the secondary. We adopted a simple model consisting of primary and secondary objects modelled with the same modified NEATM from \cite{santos-sanz_tnos_2012}. Both objects have the same geometric albedo and beaming parameter, differing only on their diameters, with the constrain that the combined brightness of both objects must yield an absolute magnitude of the system of 
$H_V=5.4$. The model has three independent parameters: The diameter of the primary $D_{\text{eq}}$, the diameter of the secondary $d_{\text{eq}}$, and the beaming parameter $\eta$. 

We implemented a Bayesian approach to derive posterior distributions similar to the occultation analysis. The likelihood has the same functional form as Eq. \ref{eq:lhood}, where $f_i$ and $\sigma _i$ are the thermal fluxes and their corresponding uncertainties, and $m(t_i|\theta)$ is the modelled flux (sum of the two bodies) at each of the $N=3$ wavelengths $t_i$ for the model parameters $\theta = (D_{\text{eq}}, d_{\text{eq}}, \eta)$. As the equivalent diameter derived from the occultation is a direct measurement, we used the posterior probability distribution in Fig. \ref{fig:DeqsComparison} (sampling 1) as the diameter of the primary object, leaving only $d_{\text{eq}}$ and $\eta$ as free parameters to explore. 

We used the TNO sample of beaming parameters from the TNOs are Cool programme \cite[Table 3]{lellouch_tnos_2013} to derive a prior distribution for the beaming parameter $\eta$, excluding centaurs to be consistent with the previous analysis (see Appendix \ref{ap:priors} for details). Since we have no a priori information about the putative satellite, we chose a uniform prior $U[0, 400]$ km for $d_\text{eq}$, excluding exploring the unphysical region of negative diameters. 

\begin{figure}
\centering
\includegraphics[width=\hsize]{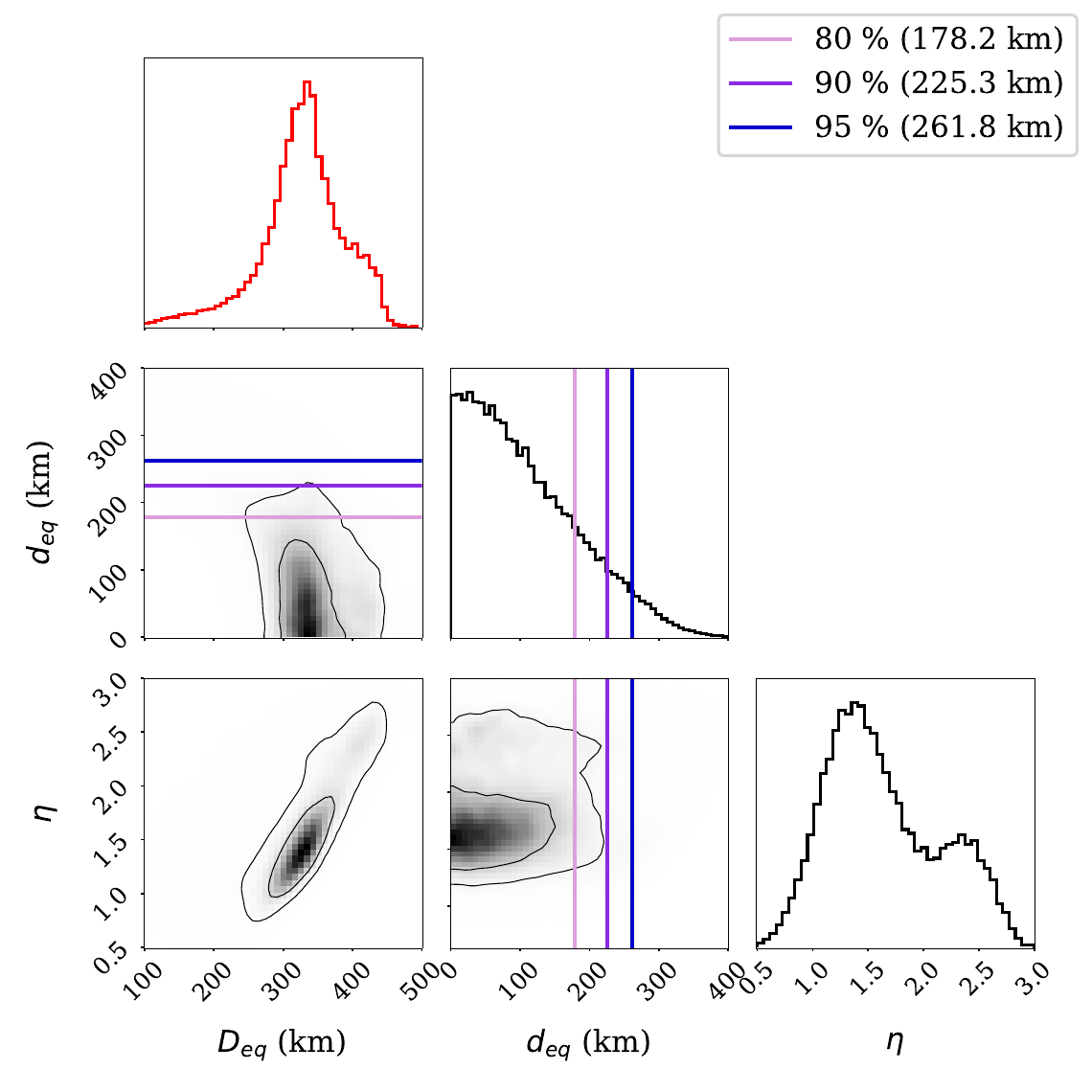}
\caption{Pairwise marginal plot for the sample of our thermal binary-object model MCMC run. The marginalised distribution in $D_{\text{eq}}$ is the same as the posterior from the occultation (sampling 1 in Fig. \ref{fig:DeqsComparison}). We show it in red to highlight that this parameter is not being updated in this sampling. The regions depicted in the 2D histograms encircle the 39.4~\% ($1\sigma$) and 78.8~\% ($2\sigma$) of the samples. The coloured lines represent quantiles of the marginalised $d_\text{eq}$ distribution. These are interpreted as upper limits for the diameter of the satellite.}
\label{fig:satellite_MCMC}
\end{figure}

Figure \ref{fig:satellite_MCMC} shows the pairwise marginal plot obtained after performing a MCMC sampling of the posterior distribution using the \texttt{emcee} package with $n_w = 256$ random walkers, $n_\text{iter} = 25000 $ iterations steps, and $n_\text{burn} = 10000$ burn in steps. We verified that the burn in phase is at least 10 times the estimated autocorrelation time to ensure convergence of the posterior distribution.

The posterior pdf reaches its peak at $d_\text{eq} = 0$ km indicating that the occultation and thermal data are better explained by a single object in our simple model. The presence of a sizeable satellite cannot be completely ruled out with the available data as indicated by the posterior of $d_\text{eq}$ having an extended wing.
In Fig. \ref{fig:satellite_MCMC} we plot the 80 \%, 90 \% and 95 \% percentiles of the marginalised distribution in $d_\text{eq}$. They represent quantitative upper limits for the size of a putative satellite. The interpretation is that, according to our sampling and model, if \objname has a satellite there is a 95 \% probability that the satellite is smaller than 261.8 km in diameter, a 90\% probability that it is smaller than 225.3 km and a 80\% that it is smaller than 178.2 km.

The lower-left panel in Fig. \ref{fig:satellite_MCMC} shows a strong correlation between $\eta$ and $D_\text{eq}$. A higher beaming parameter implies a reduced thermal emission, and hence larger objects can be allocated. The diameters of the primary and secondary are mildly correlated, as smaller primary diameters allow the presence of a larger satellite. 
We also observe that the posterior distribution of the beaming parameter shows a secondary peak at $\eta \sim 2.3$, resulting in a higher median $\eta$ than the single-body model. 
The extended wing in the joint marginal distribution of $\eta$ and $D_\text{eq}$ (Fig. \ref{fig:satellite_MCMC}, lower-left panel) indicates that the feature in the posterior of $\eta$ is produced by the distribution of $D_\text{eq}$ and the right-skewed prior in $\eta$ (see Appendix \ref{ap:priors}). The larger values in $D_\text{eq}$ allow the larger values of $\eta$ in the prior to show up in the posterior.

\section{Summary and conclusions}
On 2022 August 22, the detached trans-Neptunian object  \objnamelong caused an occultation of the \textit{Gaia} DR3 star 2727866328215869952. The event was observed from 18 sites, resulting in 13 light curves (3 lost due to bad weather, 2 discarded due to data quality). Partial flux drops of $\sim 25$ \% and $\sim 75$ \% were observed from different locations. This is a distinctive sign that two different stars have been occulted, confirming the discovery of a double star. No secondary features indicating rings, debris, or satellites were detected in the light curves.

This is the third-smallest TNO  with an occultation-derived shape to date, third only to Arrokoth \citep{buie_size_2020} and 2003~UY117 \citep{kretlow_physical_2024}. It is clear that this sample should be further enlarged if we intend to determine the critical diameter (material-dependent) that sets a limit above which TNOs adopt hydrostatic equilibrium shapes, as proposed in \cite{tancredi_which_2008}. This is not an easy task since the bulk of TNOs similar in size to \objname have $\sim 300$ mas uncertainty in their ephemerides \citep{ortiz_chapter_2020}, which translates to $\sim$ 8000 km in uncertainty for the shadow path. Consequently, we must carry out dedicated astrometric campaigns, such as those conducted in the present work, to correctly predict stellar occultations.

G!kún\textdoublepipe \textprimstress hòmdímà is the only other detached TNO with a published physical characterisation through occultations. This population is believed to have suffered a dynamical evolution similar to the scattered TNOs and substantially different to the Cold Classicals \citep{benedetti-rossi_trans-neptunian_2019}. Whether the difference in evolutionary paths is somehow reflected in the shape is still impossible to determine because of the scarcity of the sample.

An eight-parameter model was applied to the occultation light curves using a Bayesian approach, employing the \texttt{emcee} Python package to derive the posterior probability distribution. Our samplings were run for enough time to ensure convergence. A prior sensitivity analysis shows that the parameter posteriors are strongly dependent on the collected occultation data and are robust against variation in prior distributions.

From our preferred sampling we have determined a projected axis ratio of $b/a = 0.58 \pm 0.16$, which is the smallest value to date for a TNO (see Table \ref{tab:boas}). The obtained area-equivalent diameter is $D_{\text{eq}} = 330 ^{+56}_{-55}$ km and the geometric albedo $p_V = 11.2 ^{+2.1} _{-5.0}$ \%. These are compatible with the thermal data from \textit{Herschel}/PACS, available in \cite{santos-sanz_tnos_2012}. 

The present work demonstrates that stellar occultations represent a great opportunity to complement, and sometimes reinterpret, the thermal data from the TNOs are Cool programme. We were able to reduce the uncertainty on the value of the beaming parameter for \objname from $\eta = 1.17^{+1.11}_{-0.64}$ \citep{lellouch_tnos_2013} to $1.42^{+0.75}_{-0.58}$. Proceeding similarly for other TNOs are Cool objects only observed with \textit{Herschel}/PACS (poorly constrained $\eta$) would be of great interest. In this way, a more statistically significant analysis could be performed on the surface properties of TNOs from the distribution of beaming parameters and the correlation with other parameters.

The results from this paper also evidence that stellar occultations enable milliarcsecond-level astrometry for distant Solar System objects (as put forward in \citealt{rommel_stellar_2020}) and the stellar components in the case of double sources (see Table \ref{tab:astrometry}). As shown, this level of accuracy can be achieved in the case of moderate S/N light curves, even if they only scan a limited region of the body's limb (Fig. \ref{fig:chords&result}).

The fact that the occultation-derived diameter is compatible with that from thermal data is, in principle, evidence against the existence of any putative large satellite. Satellites would increase the measured thermal flux and the estimated thermal diameter would be systematically enlarged. The discussion usually ends here for past works analysing stellar occultation results for TNOs are Cool targets. We implemented a novel methodology to quantitatively set an upper limit on the size of a putative satellite for objects with thermal and occultation data.

Although our data are well explained by a single object model, the presence of a satellite of size comparable to the main body cannot be completely ruled out due to the uncertainties in the occultation size and thermal flux. We derive an upper limit of 225.3 km for a putative satellite. A better constrain in the diameter from occultations, thermal measurements with a better wavelength coverage or smaller thermal flux uncertainties would put tighter constrains in a putative satellite.
Although our analysis involves a simple model, we emphasise that simple conclusions from the compatibility of thermal and occultation diameters must be drawn with caution.

Imaging from HST could detect a satellite if the projected separation is greater than $\sim 1000$ km at \objname distance. A close-in orbit is also possible since the occultation sites have several $>250$ km gaps where the satellite could be allocated. Determining the fraction of tight binaries is of great interest since it can give us insights into key parameters of Solar System evolution, such as the lifespan of the massive disk \citep{nesvorny_binary_2019}. The methodology outlined in this paper offers a valuable approach to addressing this issue, particularly in cases involving well-constrained occultations and thermal data with good wavelength coverage and S/N.

To further refine the physical characterisation of \objname, it would be ideal to determine the rotation period and establish the rotational phase at the occultation epoch. This would help us correct for some systematic effects since the analysis of the beaming parameter assumes that the occultation and photometric measurements were taken at the same phase as the occultation. However, this assumption is reasonable since the rotational light curve amplitude of \objname is less than 0.1 magnitudes. Therefore, the rotational variability of the area-equivalent diameter is $\lesssim 5$ \%.

The determination of a rotational period would also help to constrain the three-dimensional shape, allowing us to reanalyse the thermal data with a thermophysical model to better constrain the object surface properties.

Further stellar occultations would also greatly contribute to improving \objname models. Inconveniently, \objname is currently moving in a region very sparsely populated by stars, reducing the chances of detecting future events, but still some opportunities are available for medium and large telescopes in the 2025-2030 time frame.

\section*{Data Availability}
Occultation light curve only available in electronic form at the CDS via anonymous ftp to cdsarc.u-strasbg.fr (130.79.128.5) or via \url{http://cdsweb.u-strasbg.fr/cgi-bin/qcat?J/A+A/}

\begin{acknowledgements}

The authors thank the referees and the editor for their valuable feedback. This publication is funded by the Spanish Ministry of Universities through the university training programme FPU2022/00492. The authors of IAA-CSIC acknowledge financial support from the Severo Ochoa grant CEX2021-001131-S funded by MCIN/AEI/ 10.13039/501100011033. Part of this work was supported by the Spanish projects PID2020- 112789GB-I00 from AEI and Proyecto de Excelencia de la Junta de Andalucía PY20-01309. P.S-S. acknowledges financial support by the Spanish grants PID2022-139555NB-I00
and PDC2022-133985-I00 funded by MCIN/AEI/10.13039/501100011033 and by the European Union "NextGenerationEU"/PRTR. This work is based on observations collected at the Centro Astronómico Hispano en Andalucía (CAHA) at Calar Alto, operated jointly by Junta de Andalucía and Consejo Superior de Investigaciones Científicas (IAA-CSIC). This work is based on observations made with the Liverpool Telescope operated on the island of La Palma by the Instituto de Astrofísica de Canarias in the Spanish Roque de los Muchachos Observatory. This work uses data from GAL Hassin - Centro Internazionale per le Scienze Astronomiche, Via della Fontana Mitri, 90010, Isnello (PA), Italy. IST60 telescope and its equipment are supported by the Republic of Turkey Ministry of Development (2016K12137) and Istanbul University with the project numbers BAP-3685, FBG-2017- 23943. Based on observations collected at the Observatorio de Sierra Nevada, operated by the Instituto de Astrofísica de Andalucía (IAA-CSIC). RIS acknowledges financial support from the Spanish Ministry of Science and Innovation through the project PID2022-138896NA-C54. TSR acknowledges funding from Ministerio de Ciencia e Innovación (Spanish Government), PGC2021, PID2021-125883NB-C21. FBR acknowledges CNPq grant 316604/2023-2. This work was (partially) supported by the Spanish MICIN/AEI/10.13039/501100011033 and by "ERDF A way of making Europe" by the “European Union” through grant PID2021-122842OB-C21, and the Institute of Cosmos Sciences University of Barcelona (ICCUB, Unidad de Excelencia ’María de Maeztu’) through grant CEX2019-000918-M. This work makes use of observations from the Las Cumbres Observatory global telescope network. Partially based on observations made with the Tx40 telescope at the Observatorio Astrofísico de Javalambre in Teruel, a Spanish Infraestructura Cientifico-Técnica Singular (ICTS) owned, managed, and operated by the Centro de Estudios de Física del Cosmos de Aragón (CEFCA). Tx40 is funded with the Fondos de Inversiones de Teruel (FITE). J.d.W. and MIT gratefully acknowledge financial support from the Heising-Simons Foundation, Dr. and Mrs. Colin Masson and Dr. Peter A. Gilman for Artemis, the first telescope of the SPECULOOS network situated in Tenerife, Spain.

\end{acknowledgements}

\bibliographystyle{aa}
\bibliography{2007OC10}

\begin{thebibliography}{53}
\expandafter\ifx\csname natexlab\endcsname\relax\def\natexlab#1{#1}\fi

\bibitem[{Alvarez-Candal {et~al.}(2019)Alvarez-Candal, Ayala-Loera, Gil-Hutton, Ortiz, Santos-Sanz, \& Duffard}]{alvarez-candal_absolute_2019}
Alvarez-Candal, A., Ayala-Loera, C., Gil-Hutton, R., {et~al.} 2019, MNRAS, 488, 3035

\bibitem[{Barucci \& Merlin(2020)}]{barucci_chapter_2020}
Barucci, M.~A. \& Merlin, F. 2020, in The {Trans}-{Neptunian} {Solar} {System}, ed. D.~Prialnik, M.~A. Barucci, \& L.~A. Young (Elsevier), 109--126

\bibitem[{Benedetti-Rossi {et~al.}(2019)Benedetti-Rossi, Santos-Sanz, Ortiz, Assafin, Sicardy, Morales, Vieira-Martins, Duffard, Braga-Ribas, Rommel, Camargo, Desmars, Colas, Vachier, Alvarez-Candal, Fernández-Valenzuela, Almenares, Artola, Baum, Behrend, Bérard, Bianco, Brosch, Ceretta, Colazo, Gomes-Junior, Ivanov, Jehin, Kaspi, Lecacheux, Maury, Melia, Moindrot, Morgado, Opitom, Peyrot, Pollock, Pratt, Roland, Spagnotto, Tancredi, Teng, Cacella, Emilio, Feys, Gil-Hutton, Jacques, Machado, Malacarne, Manulis, Milone, Rojas, \& Sfair}]{benedetti-rossi_trans-neptunian_2019}
Benedetti-Rossi, G., Santos-Sanz, P., Ortiz, J.~L., {et~al.} 2019, AJ, 158, 159

\bibitem[{Benedetti-Rossi {et~al.}(2016)Benedetti-Rossi, Sicardy, Buie, Ortiz, Vieira-Martins, Keller, Braga-Ribas, Camargo, Assafin, Morales, Duffard, Dias-Oliveira, Santos-Sanz, Desmars, Gomes-Júnior, Leiva, Bardecker, Bean~Jr., Olsen, Ruby, Sumner, Thirouin, Gómez-Muñoz, Gutierrez, Wasserman, Charbonneau, Irwin, Levine, \& Skiff}]{benedetti-rossi_results_2016}
Benedetti-Rossi, G., Sicardy, B., Buie, M.~W., {et~al.} 2016, AJ, 152, 156

\bibitem[{Braga-Ribas {et~al.}(2014)Braga-Ribas, Sicardy, Ortiz, Snodgrass, Roques, Vieira-Martins, Camargo, Assafin, Duffard, Jehin, Pollock, Leiva, Emilio, Machado, Colazo, Lellouch, Skottfelt, Gillon, Ligier, Maquet, Benedetti-Rossi, Gomes, Kervella, Monteiro, Sfair, Moutamid, Tancredi, Spagnotto, Maury, Morales, Gil-Hutton, Roland, Ceretta, Gu, Wang, Harpsøe, Rabus, Manfroid, Opitom, Vanzi, Mehret, Lorenzini, Schneiter, Melia, Lecacheux, Colas, Vachier, Widemann, Almenares, Sandness, Char, Perez, Lemos, Martinez, Jørgensen, Dominik, Roig, Reichart, LaCluyze, Haislip, Ivarsen, Moore, Frank, \& Lambas}]{braga-ribas_ring_2014}
Braga-Ribas, F., Sicardy, B., Ortiz, J.~L., {et~al.} 2014, Nat, 508, 72

\bibitem[{Buie \& Keller(2016)}]{buie_research_2016}
Buie, M.~W. \& Keller, J.~M. 2016, AJ, 151, 73

\bibitem[{Buie {et~al.}(2020)Buie, Porter, Tamblyn, Terrell, Parker, Baratoux, Kaire, Leiva, Verbiscer, Zangari, Colas, Diop, Samaniego, Wasserman, Benecchi, Caspi, Gwyn, Kavelaars, Ocampo~Uría, Rabassa, Skrutskie, Soto, Tanga, Young, Stern, Andersen, Arango~Pérez, Arredondo, Artola, Bâ, Ballet, Blank, Bop, Bosh, Camino~López, Carter, Castro-Chacón, Caycedo~Desprez, Caycedo~Guerra, Conard, Dauvergne, Dean, Dean, Desmars, Dieng, Bousso~Dieng, Diouf, Dorego, Dunham, Dunham, Durantini~Luca, Edwards, Erasmus, Faye, Faye, Ferrario, Ferrell, Finley, Fraser, Friedli, Galvez~Serna, Garcia-Migani, Genade, Getrost, Gil-Hutton, Gimeno, Golub, González~Murillo, Grusin, Gurovich, Hanna, Henn, Hinton, Hughes, Josephs~Jr, Joya, Kammer, Keeney, Keller, Kramer, Levine, Lisse, Lovell, Mackie, Makarchuk, Manzano, Mbaye, Mbaye, Melia, Moreno, Moss, Ndaiye, Ndiaye, Nelson, Olkin, Olsen, Ospina~Moreno, Pasachoff, Pereyra, Person, Pinzón, Pulver, Quintero, Regester, Resnick, Reyes-Ruiz, Rolfsmeier, Ruhland, Salmon,
  Santos-Sanz, Santucho, Sepúlveda~Niño, Sickafoose, Silva, Singer, Skipper, Slivan, Smith, Spagnotto, Stephens, Strabala, Tamayo, Throop, Torres~Cañas, Toure, Traore, Tsang, Turner, Vanegas, Venable, Wilson, Zuluaga, \& Zuluaga}]{buie_size_2020}
Buie, M.~W., Porter, S.~B., Tamblyn, P., {et~al.} 2020, AJ, 159, 130

\bibitem[{Castro-Ginard {et~al.}(2024)Castro-Ginard, Penoyre, Casey, Brown, Belokurov, Cantat-Gaudin, Drimmel, Fouesneau, Khanna, Kurbatov, Price-Whelan, Rix, \& Smart}]{castro2024gaia}
Castro-Ginard, A., Penoyre, Z., Casey, A.~R., {et~al.} 2024, A \& A, 688, A1

\bibitem[{Chandrasekhar(1987)}]{chandrasekhar_ellipsoidal_1987}
Chandrasekhar, S. 1987, Ellipsoidal figures of equilibrium (Dover Publications)

\bibitem[{Dias-Oliveira {et~al.}(2017)Dias-Oliveira, Sicardy, Ortiz, Braga-Ribas, Leiva, Vieira-Martins, Benedetti-Rossi, Camargo, Assafin, Gomes-Júnior, Baug, Chandrasekhar, Desmars, Duffard, Santos-Sanz, Ergang, Ganesh, Ikari, Irawati, Jain, Liying, Richichi, Shengbang, Behrend, Benkhaldoun, Brosch, Daassou, Frappa, Gal-Yam, Garcia-Lozano, Gillon, Jehin, Kaspi, Klotz, Lecacheux, Mahasena, Manfroid, Manulis, Maury, Mohan, Morales, Ofek, Rinner, Sharma, Sposetti, Tanga, Thirouin, Vachier, Widemann, Asai, Hayato, Hiroyuki, Owada, Yamamura, Hayamizu, Bradshaw, Kerr, Tomioka, Andersson, Dangl, Haymes, Naves, \& Wortmann}]{dias-oliveira_study_2017}
Dias-Oliveira, A., Sicardy, B., Ortiz, J.~L., {et~al.} 2017, AJ, 154, 22

\bibitem[{Elliot {et~al.}(1978)Elliot, Dunham, Wasserman, Millis, \& Churms}]{elliot_radii_1978}
Elliot, J.~L., Dunham, E., Wasserman, L.~H., Millis, R.~L., \& Churms, J. 1978, AJ, 83, 980

\bibitem[{Foreman-Mackey {et~al.}(2013)Foreman-Mackey, Hogg, Lang, \& Goodman}]{foreman-mackey_emcee_2013}
Foreman-Mackey, D., Hogg, D.~W., Lang, D., \& Goodman, J. 2013, PASP, 125, 306

\bibitem[{Gladman {et~al.}(2008)Gladman, Marsden, \& VanLaerhoven}]{gladman_nomenclature_2008}
Gladman, B., Marsden, B.~G., \& VanLaerhoven, C. 2008, in The {Solar} {System} beyond {Neptune}, 1st edn., ed. M.~Barucci, H.~Boehnhardt, D.~Cruikshank, \& A.~Morbidelli, Space {Science} {Series} (Tucson: University of Arizona Press), 43--57

\bibitem[{Goodman \& Weare(2010)}]{goodman_ensemble_2010}
Goodman, J. \& Weare, J. 2010, Commun. Appl. Math. Comput. Sci., 5, 65

\bibitem[{Gregory(2005)}]{gregory_bayesian_2005}
Gregory, P. 2005, Bayesian {Logical} {Data} {Analysis} for the {Physical} {Sciences}: {A} {Comparative} {Approach} with {Mathematica}® {Support} (Cambridge University Press)

\bibitem[{Harris(1998)}]{harris_thermal_1998}
Harris, A.~W. 1998, Icarus, 131, 291

\bibitem[{Karttunen {et~al.}(2017)Karttunen, Kröger, Oja, Poutanen, \& Donner}]{karttunen_solar_2017}
Karttunen, H., Kröger, P., Oja, H., Poutanen, M., \& Donner, K.~J. 2017, in Fundamental {Astronomy}, ed. H.~Karttunen, P.~Kröger, H.~Oja, M.~Poutanen, \& K.~J. Donner (Berlin, Heidelberg: Springer), 141--179

\bibitem[{Kilic {et~al.}(2022)Kilic, Braga-Ribas, Kaplan, Erece, Souami, Dindar, Desmars, Sicardy, Morgado, Shameoni, Rommel, \& Gomes-Júnior}]{kilic_occultation_2022}
Kilic, Y., Braga-Ribas, F., Kaplan, M., {et~al.} 2022, MNRAS, 515, 1346

\bibitem[{Kretlow {et~al.}(2024{\natexlab{a}})Kretlow, Ortiz, Desmars, Morales, Rommel, Santos-Sanz, Vara-Lubiano, Fernández-Valenzuela, Alvarez-Candal, Duffard, Braga-Ribas, Sicardy, Castro-Tirado, Fernández-García, Sánchez, Sota, Assafin, Benedetti-Rossi, Boufleur, Camargo, Cikota, Gomes-Junior, Gómez-Limón, Kilic, Lecacheux, Leiva, Marques-Oliveira, Morales, Morgado, Rizos, Roques, Souami, Vieira-Martins, Alarcon, Boninsegna, Çakır, Casarramona, Castellani, Cueva, Fişek, Guijarro, Haymes, Jehin, Kidd, Licandro, Maestre, Murgas, Pallé, Popescu, Pratt, Serra-Ricart, \& Talbot}]{kretlow_physical_2024}
Kretlow, M., Ortiz, J.~L., Desmars, J., {et~al.} 2024{\natexlab{a}}, A \& A, 691, A31

\bibitem[{Kretlow {et~al.}(2024{\natexlab{b}})Kretlow, Weber, Guhl, \& Tunsch}]{kretlow_cora_2024}
Kretlow, M., Weber, C., Guhl, K., \& Tunsch, E. 2024{\natexlab{b}}, Cora and {Sodis} - {Web} {Portals} and {Services} for {Stellar} {Occultation} {Work}

\bibitem[{Leiva {et~al.}(2020)Leiva, Buie, Keller, Wasserman, Kavelaars, Bridges, Haley, Strauss, Wilde, Weryk, Kervella, Baker, Bock, Conway, Cota, Estes, García, Kehrli, McCandless, McCandless, Self, Settlemire, Swanson, Thompson, \& Wise}]{leiva_stellar_2020}
Leiva, R., Buie, M.~W., Keller, J.~M., {et~al.} 2020, PSJ, 1, 48

\bibitem[{Lellouch {et~al.}(2013)Lellouch, Santos-Sanz, Lacerda, Mommert, Duffard, Ortiz, Müller, Fornasier, Stansberry, Kiss, Vilenius, Mueller, Peixinho, Moreno, Groussin, Delsanti, \& Harris}]{lellouch_tnos_2013}
Lellouch, E., Santos-Sanz, P., Lacerda, P., {et~al.} 2013, A \& A, 557, A60

\bibitem[{Mikosch {et~al.}(2006)Mikosch, Resnick, \& Robinson}]{noauthor_numerical_2006}
Mikosch, T.~V., Resnick, S.~I., \& Robinson, S.~M. 2006, Numerical {Optimization}, Springer {Series} in {Operations} {Research} and {Financial} {Engineering} (Springer New York)

\bibitem[{Morales {et~al.}(2022)Morales, Ortiz, Morales, Fernández-Valenzuela, Santos-Sanz, Duffard, Kretlow, \& Vara}]{morales_absolute_2022}
Morales, N., Ortiz, J.~L., Morales, R., {et~al.} 2022, EPSC2022

\bibitem[{Müller {et~al.}(2010)Müller, Lellouch, Stansberry, Kiss, Santos-Sanz, Vilenius, Protopapa, Moreno, Mueller, Delsanti, Duffard, Fornasier, Groussin, Harris, Henry, Horner, Lacerda, Lim, Mommert, Ortiz, Rengel, Thirouin, Trilling, Barucci, Crovisier, Doressoundiram, Dotto, Gutiérrez, Hainaut, Hartogh, Hestroffer, Kidger, Lara, Swinyard, \& Thomas}]{muller_tnos_2010}
Müller, T.~G., Lellouch, E., Stansberry, J., {et~al.} 2010, A \& A, 518, L146

\bibitem[{Nesvorný(2018)}]{nesvorny_dynamical_2018}
Nesvorný, D. 2018, ARA\&A, 56, 137

\bibitem[{Nesvorný \& Vokrouhlický(2019)}]{nesvorny_binary_2019}
Nesvorný, D. \& Vokrouhlický, D. 2019, Icarus, 331, 49

\bibitem[{Noll {et~al.}(2020)Noll, Grundy, Nesvorn{\'y}, \& Thirouin}]{Noll2020}
Noll, K.~S., Grundy, W.~M., Nesvorn{\'y}, D., \& Thirouin, A. 2020, in The {{Trans-Neptunian Solar System}}, ed. D.~Prialnik, M.~A. Barucci, \& L.~A. Young (Elsevier), 205--224

\bibitem[{Ortiz {et~al.}(2023)Ortiz, Pereira, Sicardy, Braga-Ribas, Takey, Fouad, Shaker, Kaspi, Brosch, Kretlow, Leiva, Desmars, Morgado, Morales, Vara-Lubiano, Santos-Sanz, Fernández-Valenzuela, Souami, Duffard, Rommel, Kilic, Erece, Koseoglu, Ege, Morales, Alvarez-Candal, Rizos, Gómez-Limón, Assafin, Vieira-Martins, Gomes-Júnior, Camargo, \& Lecacheux}]{ortiz_changing_2023}
Ortiz, J.~L., Pereira, C.~L., Sicardy, B., {et~al.} 2023, A \& A, 676, L12

\bibitem[{Ortiz {et~al.}(2017)Ortiz, Santos-Sanz, Sicardy, Benedetti-Rossi, Bérard, Morales, Duffard, Braga-Ribas, Hopp, Ries, Nascimbeni, Marzari, Granata, Pál, Kiss, Pribulla, Komžík, Hornoch, Pravec, Bacci, Maestripieri, Nerli, Mazzei, Bachini, Martinelli, Succi, Ciabattari, Mikuz, Carbognani, Gaehrken, Mottola, Hellmich, Rommel, Fernández-Valenzuela, Bagatin, Cikota, Cikota, Lecacheux, Vieira-Martins, Camargo, Assafin, Colas, Behrend, Desmars, Meza, Alvarez-Candal, Beisker, Gomes-Junior, Morgado, Roques, Vachier, Berthier, Mueller, Madiedo, Unsalan, Sonbas, Karaman, Erece, Koseoglu, Ozisik, Kalkan, Guney, Niaei, Satir, Yesilyaprak, Puskullu, Kabas, Demircan, Alikakos, Charmandaris, Leto, Ohlert, Christille, Szakáts, Farkas, Varga-Verebélyi, Marton, Marciniak, Bartczak, Santana-Ros, Butkiewicz-Bąk, Dudziński, Alí-Lagoa, Gazeas, Tzouganatos, Paschalis, Tsamis, Sánchez-Lavega, Pérez-Hoyos, Hueso, Guirado, Peris, \& Iglesias-Marzoa}]{ortiz_size_2017}
Ortiz, J.~L., Santos-Sanz, P., Sicardy, B., {et~al.} 2017, Nat, 550, 219

\bibitem[{Ortiz {et~al.}(2020{\natexlab{a}})Ortiz, Santos-Sanz, Sicardy, Benedetti-Rossi, Duffard, Morales, Braga-Ribas, Fernández-Valenzuela, Nascimbeni, Nardiello, Carbognani, Buzzi, Aletti, Bacci, Maestripieri, Mazzei, Mikuz, Skvarc, Ciabattari, Lavalade, Scarfi, Mari, Conjat, Sposetti, Bachini, Succi, Mancini, Alighieri, Dal~Canto, Masucci, Vara-Lubiano, Gutiérrez, Desmars, Lecacheux, Vieira-Martins, Camargo, Assafin, Colas, Beisker, Behrend, Mueller, Meza, Gomes-Junior, Roques, Vachier, Mottola, Hellmich, Campo~Bagatin, Alvarez-Candal, Cikota, Cikota, Christille, Pál, Kiss, Pribulla, Komžík, Madiedo, Charmandaris, Alikakos, Szakáts, Farkas-Takács, Varga-Verebélyi, Marton, Marciniak, Bartczak, Butkiewicz-B{\textbackslash}c~ak, Dudziński, Alí-Lagoa, Gazeas, Paschalis, Tsamis, Guirado, Peris, Iglesias-Marzoa, Schnabel, Manzano, Navarro, Perelló, Vecchione, Noschese, \& Morrone}]{ortiz_large_2020}
Ortiz, J.~L., Santos-Sanz, P., Sicardy, B., {et~al.} 2020{\natexlab{a}}, A \& A, 639, A134

\bibitem[{Ortiz {et~al.}(2012)Ortiz, Sicardy, Braga-Ribas, Alvarez-Candal, Lellouch, Duffard, Pinilla-Alonso, Ivanov, Littlefair, Camargo, Assafin, Unda-Sanzana, Jehin, Morales, Tancredi, Gil-Hutton, de~la Cueva, Colque, Da~Silva~Neto, Manfroid, Thirouin, Gutiérrez, Lecacheux, Gillon, Maury, Colas, Licandro, Mueller, Jacques, Weaver, Milone, Salvo, Bruzzone, Organero, Behrend, Roland, Vieira-Martins, Widemann, Roques, Santos-Sanz, Hestroffer, Dhillon, Marsh, Harlingten, Bagatin, Alonso, Ortiz, Colazo, Lima, Oliveira, Kerber, Smiljanic, Pimentel, Giacchini, Cacella, \& Emilio}]{ortiz_albedo_2012}
Ortiz, J.~L., Sicardy, B., Braga-Ribas, F., {et~al.} 2012, Nat, 491, 566

\bibitem[{Ortiz {et~al.}(2020{\natexlab{b}})Ortiz, Sicardy, Camargo, Santos-Sanz, \& Braga-Ribas}]{ortiz_chapter_2020}
Ortiz, J.~L., Sicardy, B., Camargo, J. I.~B., Santos-Sanz, P., \& Braga-Ribas, F. 2020{\natexlab{b}}, in The {Trans}-{Neptunian} {Solar} {System}, ed. D.~Prialnik, M.~A. Barucci, \& L.~A. Young (Elsevier), 413--437

\bibitem[{Pereira {et~al.}(2023)Pereira, Sicardy, Morgado, Braga-Ribas, Fernández-Valenzuela, Souami, Holler, Boufleur, Margoti, Assafin, Ortiz, Santos-Sanz, Epinat, Kervella, Desmars, Vieira-Martins, Kilic, Júnior, Camargo, Emilio, Vara-Lubiano, Kretlow, Albert, Alcock, Ball, Bender, Buie, Butterfield, Camarca, Castro-Chacón, Dunford, Fisher, Gamble, Geary, Gnilka, Green, Hartman, Huang, Januszewski, Johnston, Kagitani, Kamin, Kavelaars, Keller, Kleer, Lehner, Luken, Marchis, Marlin, McGregor, Nikitin, Nolthenius, Patrick, Redfield, Rengstorf, Reyes-Ruiz, Seccull, Skrutskie, Smith, Sproul, Stephens, Szentgyorgyi, Sánchez-Sanjuán, Tatsumi, Verbiscer, Wang, Yoshida, Young, \& Zhang}]{pereira_two_2023}
Pereira, C.~L., Sicardy, B., Morgado, B.~E., {et~al.} 2023, A \& A, 673, L4

\bibitem[{Perna {et~al.}(2013)Perna, Dotto, Barucci, Epifani, Vilenius, Dall’Ora, Fornasier, \& Müller}]{perna_photometry_2013}
Perna, D., Dotto, E., Barucci, M.~A., {et~al.} 2013, A \& A, 554, A49

\bibitem[{Poglitsch {et~al.}(2010)Poglitsch, Waelkens, Geis, Feuchtgruber, Vandenbussche, Rodriguez, Krause, Renotte, van Hoof, Saraceno, Cepa, Kerschbaum, Agn{\`e}se, Ali, Altieri, Andreani, Augueres, Balog, Barl, Bauer, Belbachir, Benedettini, Billot, Boulade, Bischof, Blommaert, Callut, Cara, Cerulli, Cesarsky, Contursi, Creten, Meester, Doublier, Doumayrou, Duband, Exter, Genzel, Gillis, Gr{\"o}zinger, Henning, Herreros, Huygen, Inguscio, Jakob, Jamar, Jean, de~Jong, Katterloher, Kiss, Klaas, Lemke, Lutz, Madden, Marquet, Martignac, Mazy, Merken, Montfort, Morbidelli, M{\"u}ller, Nielbock, Okumura, Orfei, Ottensamer, Pezzuto, Popesso, Putzeys, Regibo, Reveret, Royer, Sauvage, Schreiber, Stegmaier, Schmitt, Schubert, Sturm, Thiel, Tofani, Vavrek, Wetzstein, Wieprecht, \& Wiezorrek}]{Poglitsch2010}
Poglitsch, A., Waelkens, C., Geis, N., {et~al.} 2010, A \& A, 518, L2

\bibitem[{Rambaux {et~al.}(2017)Rambaux, Baguet, Chambat, \& Castillo-Rogez}]{rambaux_equilibrium_2017}
Rambaux, N., Baguet, D., Chambat, F., \& Castillo-Rogez, J.~C. 2017, ApJL, 850, L9

\bibitem[{Riello {et~al.}(2021)Riello, De~Angeli, Evans, Montegriffo, Carrasco, Busso, Palaversa, Burgess, Diener, Davidson, Rowell, Fabricius, Jordi, Bellazzini, Pancino, Harrison, Cacciari, van Leeuwen, Hambly, Hodgkin, Osborne, Altavilla, Barstow, Brown, Castellani, Cowell, De~Luise, Gilmore, Giuffrida, Hidalgo, Holland, Marinoni, Pagani, Piersimoni, Pulone, Ragaini, Rainer, Richards, Sanna, Walton, Weiler, \& Yoldas}]{riello_gaia_2021}
Riello, M., De~Angeli, F., Evans, D.~W., {et~al.} 2021, A \& A, 649, A3

\bibitem[{Rommel {et~al.}(2020)Rommel, Braga-Ribas, Desmars, Camargo, Ortiz, Sicardy, Vieira-Martins, Assafin, Santos-Sanz, Duffard, Fernández-Valenzuela, Lecacheux, Morgado, Benedetti-Rossi, Gomes-Júnior, Pereira, Herald, Hanna, Bradshaw, Morales, Brimacombe, Burtovoi, Carruthers, de~Barros, Fiori, Gilmore, Hooper, Hornoch, Jacques, Janik, Kerr, Kilmartin, Winkel, Naletto, Nardiello, Nascimbeni, Newman, Ossola, Pál, Pimentel, Pravec, Sposetti, Stechina, Szakáts, Ueno, Zampieri, Broughton, Dunham, Dunham, Gault, Hayamizu, Hosoi, Jehin, Jones, Kitazaki, Komžík, Marciniak, Maury, Mikuž, Nosworthy, Fábrega~Polleri, Rahvar, Sfair, Siqueira, Snodgrass, Sogorb, Tomioka, Tregloan-Reed, \& Winter}]{rommel_stellar_2020}
Rommel, F.~L., Braga-Ribas, F., Desmars, J., {et~al.} 2020, A \& A, 644, A40

\bibitem[{Rommel {et~al.}(2023)Rommel, Braga-Ribas, Ortiz, Sicardy, Santos-Sanz, Desmars, Camargo, Vieira-Martins, Assafin, Morgado, Boufleur, Benedetti-Rossi, Gomes-Júnior, Fernández-Valenzuela, Holler, Souami, Duffard, Margoti, Vara-Lubiano, Lecacheux, Plouvier, Morales, Maury, Fabrega, Ceravolo, Jehin, Albanese, Mariey, Cikota, Ruždjak, Cikota, Szakáts, Aissa, Gringahcene, Kashuba, Koshkin, Zhukov, Fişek, Çakir, Özer, Schnabel, Schnabel, Signoret, Morrone, Santana-Ros, Pereira, Emilio, Burdanov, Wit, Barkaoui, Gillon, Leto, Frasca, Catanzaro, Sanchez, Tagliaferri, Sora, Isopi, Krugly, Slyusarev, Chiorny, Mikuž, Bacci, Maestripieri, Grazia, Cueva, Yuste-Moreno, Ciabattari, Kozhukhov, Serra-Ricart, Alarcon, Licandro, Masi, Bacci, Bosch, Behem, Prost, Renner, Conjat, Bachini, Succi, Stoian, Juravle, Carosati, Gowe, Carrillo, Zheleznyak, Montigiani, Foster, Mannucci, Ruocco, Cuevas, Marcantonio, Coretti, Iafrate, Baldini, Collins, Pál, Csák, Fernández-Garcia, Castro-Tirado, Hudin, Madiedo, Anghel,
  Calvo-Fernández, Valvasori, Guido, Gherase, Kamoun, Fafet, Sánchez-González, Curelaru, Vîntdevară, Danescu, Gout, Schmitz, Sota, Belskaya, Rodríguez-Marco, Kilic, Frappa, Klotz, Lavayssière, Oliveira, Popescu, Mammana, Fernández-Lajús, Schmidt, Hopp, Komžík, Pribulla, Tomko, Husárik, Erece, Eryilmaz, Buzzi, Gährken, Nardiello, Hornoch, Sonbas, Er, Burwitz, Sybilski, Bykowski, Müller, Ogloza, Gonçalves, Ferreira, Ferreira, Bento, Meister, Bagiran, Tekeş, Marciniak, Moravec, Delinčák, Gianni, Casalnuovo, Boutet, Sanchez, Klemt, Wuensche, Burzynski, Borkowski, Serrau, Dangl, Klös, Weber, Urbaník, Rousselot, Kubánek, André, Colazo, Spagnotto, Sickafoose, Hueso, Sánchez-Lavega, Fisher, Rengstorf, Perelló, Dascalu, Altan, Gazeas, Santana, Sfair, Winter, Kalkan, Canales-Moreno, Trigo-Rodríguez, Tsamis, Tigani, Sioulas, Lekkas, Bertesteanu, Dumitrescu, Wilberger, Barnes, Fieber-Beyer, Swaney, Fuentes, Mendez, Dumitru, Flynn, \& Wake}]{rommel_large_2023}
Rommel, F.~L., Braga-Ribas, F., Ortiz, J.~L., {et~al.} 2023, A \& A, 678, A167

\bibitem[{Santos-Sanz {et~al.}(2012)Santos-Sanz, Lellouch, Fornasier, Kiss, Pal, Müller, Vilenius, Stansberry, Mommert, Delsanti, Mueller, Peixinho, Henry, Ortiz, Thirouin, Protopapa, Duffard, Szalai, Lim, Ejeta, Hartogh, Harris, \& Rengel}]{santos-sanz_tnos_2012}
Santos-Sanz, P., Lellouch, E., Fornasier, S., {et~al.} 2012, A \& A, 541, A92

\bibitem[{Santos-Sanz {et~al.}(2021)Santos-Sanz, Ortiz, Sicardy, Benedetti-Rossi, Morales, Fernández-Valenzuela, Duffard, Iglesias-Marzoa, Lamadrid, Maícas, Pérez, Gazeas, Guirado, Peris, Ballesteros, Organero, Ana-Hernández, Fonseca, Alvarez-Candal, Jiménez-Teja, Vara-Lubiano, Braga-Ribas, Camargo, Desmars, Assafin, Vieira-Martins, Alikakos, Boutet, Bretton, Carbognani, Charmandaris, Ciabattari, Delincak, Fuambuena~Leiva, González, Haymes, Hellmich, Horbowicz, Jennings, Kattentidt, Kiss, Komžík, Lecacheux, Marciniak, Moindrot, Mottola, Pal, Paschalis, Pastor, Perello, Pribulla, Ratinaud, Reyes, Sanchez, Schnabel, Selva, Signoret, Sonbas, \& Alí-Lagoa}]{santos-sanz_2017_2021}
Santos-Sanz, P., Ortiz, J.~L., Sicardy, B., {et~al.} 2021, MNRAS, 501, 6062

\bibitem[{Santos-Sanz {et~al.}(2022)Santos-Sanz, Ortiz, Sicardy, Popescu, Benedetti-Rossi, Morales, Vara-Lubiano, Camargo, Pereira, Rommel, Assafin, Desmars, Braga-Ribas, Duffard, Oliveira, Vieira-Martins, Fernández-Valenzuela, Morgado, Acar, Anghel, Atalay, Ateş, Bakiş, Bakis, Eker, Erece, Kaspi, Kayhan, Kilic, Kilic, Manulis, Nedelcu, Niaei, Nir, Ofek, Ozisik, Petrescu, Satir, Solmaz, Sonka, Tekes, Unsalan, Yesilyaprak, Anghel, Berteşteanu, Curelaru, Danescu, Dumitrescu, Gherase, Hudin, Stoian, Tercu, Truta, Turcu, Vantdevara, Belskaya, Dementiev, Gazeas, Karampotsiou, Kashuba, Kiss, Koshkin, Kozhukhov, Krugly, Lecacheux, Pal, Püsküllü, Szakats, Zhukov, Bamberger, Mondon, Perelló, Pratt, Schnabel, Selva, Teng, Tigani, Tsamis, Weber, Wells, Kalkan, Kudak, Marciniak, Ogloza, Özdemir, Pakštiene, Perig, \& Żejmo}]{santos-sanz_physical_2022}
Santos-Sanz, P., Ortiz, J.~L., Sicardy, B., {et~al.} 2022, A \& A, 664, A130, publisher: EDP Sciences

\bibitem[{Scott(2015)}]{scott2015multivariate}
Scott, D.~W. 2015, Multivariate density estimation: theory, practice, and visualization (John Wiley \& Sons)

\bibitem[{Souami {et~al.}(2020)Souami, Braga-Ribas, Sicardy, Morgado, Ortiz, Desmars, Camargo, Vachier, Berthier, Carry, Anderson, Showers, Thomason, Maley, Thomas, Buie, Leiva, Keller, Vieira-Martins, Assafin, Santos-Sanz, Morales, Duffard, Benedetti-Rossi, Gomes-Júnior, Boufleur, Pereira, Margoti, Pavlov, George, Oesper, Bardecker, Dunford, Kehrli, Spencer, Cota, Garcia, Lara, McCandless, Self, Lecacheux, Frappa, Dunham, \& Emilio}]{souami_multi-chord_2020}
Souami, D., Braga-Ribas, F., Sicardy, B., {et~al.} 2020, A \& A, 643, A125

\bibitem[{Stansberry {et~al.}(2008)Stansberry, Grundy, Brown, Cruikshank, Spencer, Trilling, \& Margot}]{stansberry_physical_2008}
Stansberry, J., Grundy, W., Brown, M., {et~al.} 2008, in The {Solar} {System} {Beyond} {Neptune}, ed. M.~A. Barucci, H.~Boehnhardt, D.~P. Cruikshank, A.~Morbidelli, \& R.~Dotson (The University of Arizona Press), 161--179

\bibitem[{Strauss {et~al.}(2021)Strauss, Leiva, Keller, Wilde, Buie, Weryk, Kavelaars, Bridges, Wasserman, Trilling, Ainsworth, Anthony, Baker, Bardecker, Bean, Bock, Chase, Dean, Frei, George, Gill, Gimple, Givot, Hopfe, Cota, Kehrli, King, Haley, Lara, Lund, Mattes, McCandless, McCrystal, McRae, Melgarejo, Mendoza, Miller, Norfolk, Palmquist, Reaves, Rivard, von Schalscha, Schar, Stoffel, Swanson, Thompson, Wise, Woods, \& Yang}]{strauss_sizes_2021}
Strauss, R.~H., Leiva, R., Keller, J.~M., {et~al.} 2021, PSJ, 2, 22

\bibitem[{Tancredi \& Favre(2008)}]{tancredi_which_2008}
Tancredi, G. \& Favre, S. 2008, Icarus, 195, 851

\bibitem[{Vallenari {et~al.}(2023)Vallenari, Brown, Prusti, Bruijne, Arenou, Babusiaux, Biermann, Creevey, Ducourant, Evans, Eyer, Guerra, Hutton, Jordi, Klioner, Lammers, Lindegren, Luri, Mignard, Panem, Pourbaix, Randich, Sartoretti, Soubiran, Tanga, Walton, Bailer-Jones, Bastian, Drimmel, Jansen, Katz, Lattanzi, Leeuwen, Bakker, Cacciari, Castañeda, Angeli, Fabricius, Fouesneau, Frémat, Galluccio, Guerrier, Heiter, Masana, Messineo, Mowlavi, Nicolas, Nienartowicz, Pailler, Panuzzo, Riclet, Roux, Seabroke, Sordo, Thévenin, Gracia-Abril, Portell, Teyssier, Altmann, Andrae, Audard, Bellas-Velidis, Benson, Berthier, Blomme, Burgess, Busonero, Busso, Cánovas, Carry, Cellino, Cheek, Clementini, Damerdji, Davidson, Teodoro, Campos, Delchambre, Dell’Oro, Esquej, Fernández-Hernández, Fraile, Garabato, García-Lario, Gosset, Haigron, Halbwachs, Hambly, Harrison, Hernández, Hestroffer, Hodgkin, Holl, Janßen, Fombelle, Jordan, Krone-Martins, Lanzafame, Löffler, Marchal, Marrese, Moitinho, Muinonen, Osborne,
  Pancino, Pauwels, Recio-Blanco, Reylé, Riello, Rimoldini, Roegiers, Rybizki, Sarro, Siopis, Smith, Sozzetti, Utrilla, Leeuwen, Abbas, Ábrahám, Aramburu, Aerts, Aguado, Ajaj, Aldea-Montero, Altavilla, Álvarez, Alves, Anders, Anderson, Varela, Antoja, Baines, Baker, Balaguer-Núñez, Balbinot, Balog, Barache, Barbato, Barros, Barstow, Bartolomé, Bassilana, Bauchet, Becciani, Bellazzini, Berihuete, Bernet, Bertone, Bianchi, Binnenfeld, Blanco-Cuaresma, Blazere, Boch, Bombrun, Bossini, Bouquillon, Bragaglia, Bramante, Breedt, Bressan, Brouillet, Brugaletta, Bucciarelli, Burlacu, Butkevich, Buzzi, Caffau, Cancelliere, Cantat-Gaudin, Carballo, Carlucci, Carnerero, Carrasco, Casamiquela, Castellani, Castro-Ginard, Chaoul, Charlot, Chemin, Chiaramida, Chiavassa, Chornay, Comoretto, Contursi, Cooper, Cornez, Cowell, Crifo, Cropper, Crosta, Crowley, Dafonte, Dapergolas, David, David, Laverny, Luise, March, Ridder, Souza, Torres, Peloso, Pozo, Delbo, Delgado, Delisle, Demouchy, Dharmawardena, Matteo, Diakite,
  Diener, Distefano, Dolding, Edvardsson, Enke, Fabre, Fabrizio, Faigler, Fedorets, Fernique, Fienga, Figueras, Fournier, Fouron, Fragkoudi, Gai, Garcia-Gutierrez, Garcia-Reinaldos, García-Torres, Garofalo, Gavel, Gavras, Gerlach, Geyer, Giacobbe, Gilmore, Girona, Giuffrida, Gomel, Gomez, González-Núñez, González-Santamaría, González-Vidal, Granvik, Guillout, Guiraud, Gutiérrez-Sánchez, Guy, Hatzidimitriou, Hauser, Haywood, Helmer, Helmi, Sarmiento, Hidalgo, Hilger, Hładczuk, Hobbs, Holland, Huckle, Jardine, Jasniewicz, Piccolo, Jiménez-Arranz, Jorissen, Campillo, Julbe, Karbevska, Kervella, Khanna, Kontizas, Kordopatis, Korn, Kóspál, Kostrzewa-Rutkowska, Kruszyńska, Kun, Laizeau, Lambert, Lanza, Lasne, Campion, Lebreton, Lebzelter, Leccia, Leclerc, Lecoeur-Taibi, Liao, Licata, Lindstrøm, Lister, Livanou, Lobel, Lorca, Loup, Pardo, Romeo, Managau, Mann, Manteiga, Marchant, Marconi, Marcos, Santos, Pina, Marinoni, Marocco, Marshall, Polo, Martín-Fleitas, Marton, Mary, Masip, Massari,
  Mastrobuono-Battisti, Mazeh, McMillan, Messina, Michalik, Millar, Mints, Molina, Molinaro, Molnár, Monari, Monguió, Montegriffo, Montero, Mor, Mora, Morbidelli, Morel, Morris, Muraveva, Murphy, Musella, Nagy, Noval, Ocaña, Ogden, Ordenovic, Osinde, Pagani, Pagano, Palaversa, Palicio, Pallas-Quintela, Panahi, Payne-Wardenaar, Esteller, Penttilä, Pichon, Piersimoni, Pineau, Plachy, Plum, Poggio, Prša, Pulone, Racero, Ragaini, Rainer, Raiteri, Rambaux, Ramos, Ramos-Lerate, Fiorentin, Regibo, Richards, Diaz, Ripepi, Riva, Rix, Rixon, Robichon, Robin, Robin, Roelens, Rogues, Rohrbasser, Romero-Gómez, Rowell, Royer, Mieres, Rybicki, Sadowski, Núñez, Sellés, Sahlmann, Salguero, Samaras, Gimenez, Sanna, Santoveña, Sarasso, Schultheis, Sciacca, Segol, Segovia, Ségransan, Semeux, Shahaf, Siddiqui, Siebert, Siltala, Silvelo, Slezak, Slezak, Smart, Snaith, Solano, Solitro, Souami, Souchay, Spagna, Spina, Spoto, Steele, Steidelmüller, Stephenson, Süveges, Surdej, Szabados, Szegedi-Elek, Taris, Taylor,
  Teixeira, Tolomei, Tonello, Torra, Torra, Elipe, Trabucchi, Tsounis, Turon, Ulla, Unger, Vaillant, Dillen, Reeven, Vanel, Vecchiato, Viala, Vicente, Voutsinas, Weiler, Wevers, Wyrzykowski, Yoldas, Yvard, Zhao, Zorec, Zucker, \& Zwitter}]{vallenari_gaia_2023}
Vallenari, A., Brown, A. G.~A., Prusti, T., {et~al.} 2023, A \& A, 674, A1

\bibitem[{van Belle(1999)}]{van_belle_predicting_1999}
van Belle, G.~T. 1999, PASP, 111, 1515

\bibitem[{Vara-Lubiano {et~al.}(2022)Vara-Lubiano, Benedetti-Rossi, Santos-Sanz, Ortiz, Sicardy, Popescu, Morales, Rommel, Morgado, Pereira, Álvarez Candal, Fernández-Valenzuela, Souami, Ilic, Vince, Bachev, Semkov, Nedelcu, Şonka, Hudin, Boaca, Inceu, Curelaru, Gherase, Turcu, Moldovan, Mircea, Predatu, Teodorescu, Stoian, Juravle, Braga-Ribas, Desmars, Duffard, Lecacheux, Camargo, Assafin, Vieira-Martins, Pribulla, Husárik, Sivanič, Pal, Szakats, Kiss, Alonso-Santiago, Frasca, Szabó, Derekas, Szigeti, Drozdz, Ogloza, Skvarč, Ciabattari, Delincak, Di~Marcantonio, Iafrate, Coretti, Baldini, Baruffetti, Klös, Dumitrescu, Mikuž, \& Mohar}]{vara-lubiano_multichord_2022}
Vara-Lubiano, M., Benedetti-Rossi, G., Santos-Sanz, P., {et~al.} 2022, A \& A, 663, A121

\bibitem[{Volk \& Laerhoven(2024)}]{volk_dynamical_2024}
Volk, K. \& Laerhoven, C.~V. 2024, RNAAS, 8, 36

\bibitem[{{Zacharias} {et~al.}(2004){Zacharias}, {Monet}, {Levine}, {Urban}, {Gaume}, \& {Wycoff}}]{zacharias_naval_2004}
{Zacharias}, N., {Monet}, D.~G., {Levine}, S.~E., {et~al.} 2004, in AAS Meeting Abstracts, Vol. 205, AAS Meeting Abstracts, 48.15

\end{thebibliography}

\begin{appendix} 

\section{Prior distributions}
\label{ap:priors}
\subsection{Occultation sampling: $b/a$}
\begin{figure}[h!]
\centering
\includegraphics[width=\hsize]{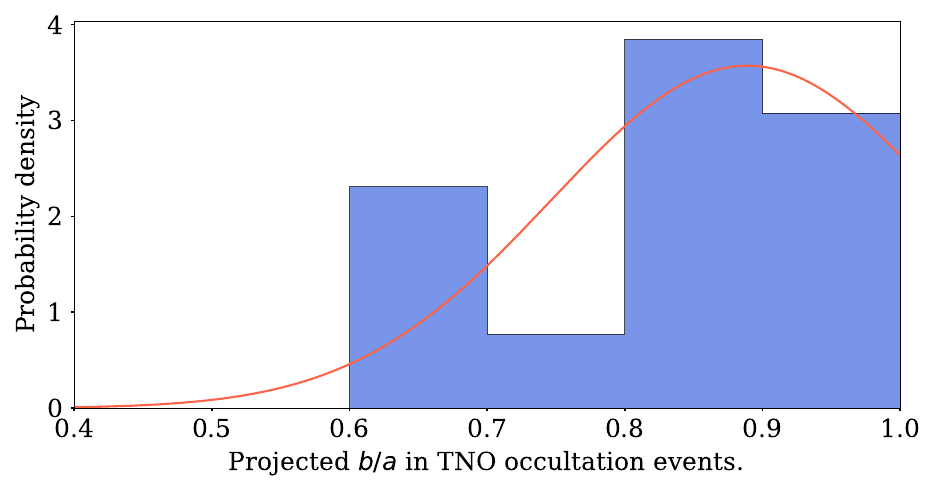}
\caption{Histogram of the available sample of projected axis ratios in published TNO occultations together with our fitted truncated normal. The latter is the used prior distribution in $b/a$ in sampling 2.}
\label{fig:boas&fit}
\end{figure}

To fit a truncated normal distribution to the $b/a$ available data points in Table \ref{tab:boas}
, we performed a maximum likelihood estimation. To find the point ($\mu$,$\sigma$) where this maximum is achieved, we implemented the Broyden, Fletcher, Goldfarb, and Shannon (BFGS) method \citep{noauthor_numerical_2006} with the mean and standard deviation of the sample as initial points. The resulting optimal truncated normal has parameters $\mu = 0.889$ and $\sigma = 0.143$. It is plotted in Fig. \ref{fig:boas&fit} together with a histogram of the sample of $b/a$ values.

Then, to obtain a sample of $D_{\text{eq}}$ values from this $b/a$ prior combined with our $N(\mu = 169\ \text{km}, \sigma = 17\ \text{km})$ prior in $a$, we extracted a 1000 size sample from both prior distributions and combine them through $D_{\text{eq}} = 2a^2(b/a)$. The result is a 1000 size sample of diameters which is used in the K-S test against the distribution from the thermal data.

\subsection{Thermal data sampling: $\eta$}
\begin{figure}[h!]
\centering
\includegraphics[width=\hsize]{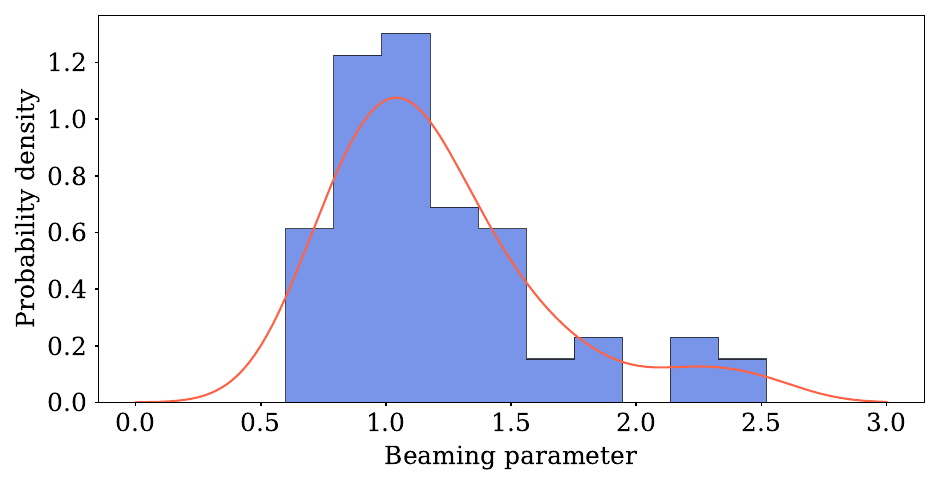}
\caption{Histogram of the available sample of beaming parameters from TNOs are Cool \citep[Centaurs excluded;][]{lellouch_tnos_2013} together with our Gaussian kernel density estimated pdf. The latter is the used prior distribution in $\eta$ in the sampling in Sect. \ref{sec:satellite}.}
\label{fig:etas&fit}
\end{figure}

In this case, as it is clear form the histogram in Fig. \ref{fig:etas&fit}, a normal distribution is not appropriate. Therefore, we implemented a Gaussian Kernel Density Estimation with the \texttt{gaussian\_kde} class of \texttt{scipy.stats} Python package. For the bandwidth selection, we implemented Scott's rule \citep{scott2015multivariate}. The resulting prior pdf for $\eta$ is shown in Fig. \ref{fig:etas&fit}

\section{Projected area change between occultation and thermal measurement epoch}
\label{ap:area_change}

To statistically estimate the change in projected area between the mean epoch of the radiometric measurements with \textit{Herschel}/PACS, 2010 October 17, and the occultation data, 2022 August 22, we considered an ellipsoidal three-dimensional shape for \objname. We extracted 100000 random ellipsoids and rotation poles. The rotation poles are extracted from an isotropic distribution on the sphere. The axis ratios of the ellipsoid $a/b$ and $c/a$ are randomly selected from uniform distributions of $U[0.5, 1]$ and $U[0.3, b/a]$ respectively, excluding only very elongated shapes from our analysis. The sampled shapes and rotation poles were then filtered. We exclude those projecting an ellipse incompatible with an axis ratio $0.58\pm 0.16$ on 2022 August 22, as observed in the occultation. We also exclude those that, on this same epoch, show a rotational light curve amplitude larger than 0.1 mag (considering flux proportional to the projected area). For those ellipsoids that are compatible with our observations, we then extracted a random rotational phase to compute the projected area on 2010 October 17, $A_{\text{thermal}}$. This area can be compared with that projected in the occultation $A_{\text{occ}}$. 

\begin{figure}[h!]
\centering
\includegraphics[width=\hsize]{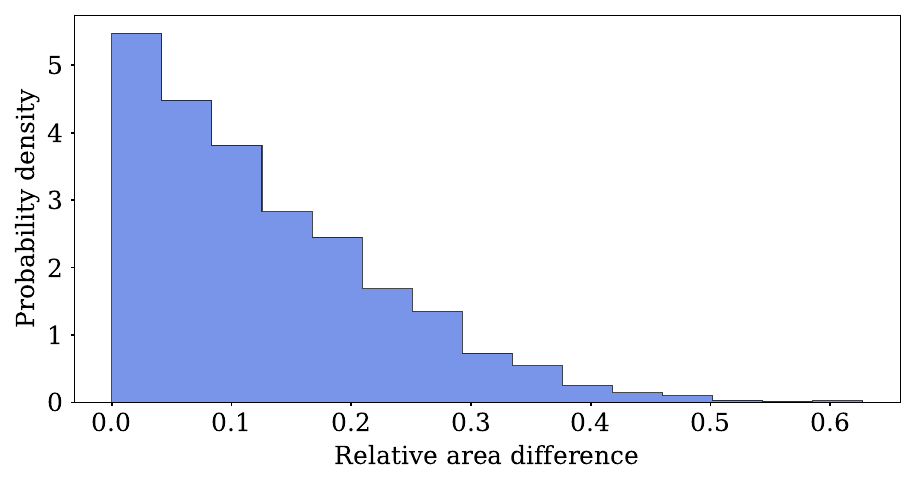}
\caption{Histogram of the relative area difference ($|A_{\text{thermal}}-A_{\text{occ}}|/A_{\text{occ}}$) for our random sample of ellipsoids and rotation poles. The epoch for $A_{\text{thermal}}$ is 2010 October 17 and the rotational phase is random. The epoch for $A_{\text{occ}}$ is 2022 August 22 and the rotational phase is such that a projected axis ratio compatible with the occultation is achieved.}
\label{fig:area_diff}
\end{figure}

In Fig. \ref{fig:area_diff}, the resulting distribution of area differences (relative to $A_{\text{occ}}$) is shown. According to this analysis, there is a 50 \% probability that the area difference is less than 10 \%, and a 86 \% probability that the area difference is less than 25 \%. However, large area differences are plausible.

\section{Observing sites data}

\begin{table*}[t]
\centering
\begin{tabular}{p{2.9cm}p{1.2cm}p{1.1cm}p{1.7cm}p{1.1cm}p{3cm}p{2.8cm}@{}}
\toprule
SiteID & UT Start  & t$_{exp}$ (s)   & Lat (deg) & Ap. (m) & Observers & Comments\\
       & UT End    & t$_{cycle}$ (s) & Lon (deg) & S/N      &           &         \\
       &           &                 & Alt (m)   &          &           &         \\
\midrule
Observatorio de Aras de los Olmos (OAO) & ... ... & ... ... & 39.945123 -1.101080 1379 & ... ... & A. Jiménez-Guisado, B. Samper-Doménech & Data not useful\\
LCO, Tenerife. Aqawan A \#1 & ... ... & ... ... & 28.300390 -16.511700 2443 & ... ... & T. Santana-Ros & Data not useful\\
Albox & 00:19:47 02:05:09 & 3.0 3.193 & 37.405575 -2.151800 538 & 0.406 18 & J. Maestre & \\
Botorrita & 01:26:02 01:36:02 & 0.5 0.5007 & 41.497375 -1.020867 456 & 0.508 10 & O. Canales Moreno, D. Lafuente Aznar, S. Calavia Belloc & Synchronization problems between the GPS and the acquisition software.\\
El Guijo Observatory & 01:30:21 01:35:20 & 1.0 1.002 & 40.607537 -4.031010 968 & 0.300 13 & A. San Segundo & \\
Guirguillano Observatory & 01:29:07 01:38:20 & 10.24 10.24 & 42.711764 -1.865110 645 & 0.310 14 & P. Martorell & Video data.\\
Istanbul University Observatory, IST60 (Çanakkale) & 01:27:45 01:55:23 & 3.0 3.726 & 40.098990 26.474490 437 & 0.600 11 & S. Fişek, S. Aliş & Timestamps truncated to the nearest second.\\
La Palma-Liverpool Telescope & 01:28:09 01:42:24 & 1.183 1.223 & 28.762574 -17.879200 2333 & 2.000 21 & J. Ortiz, R. Duffard, N. Morales & \\
Observatorio Astrofísico de Javalambre Tx40 & 01:09:13 01:45:31 & 4.0 4.444 & 40.042464 -1.016482 1988 & 0.400 18 & R. Iglesias-Marzoa, R. Infante-Sainz, T. Kuutma & \\
Oukaïmeden Observatory & 00:21:28 04:31:23 & 2.0 $\ \ $ 5.0 & 31.206302 -7.866467 2779 & 0.500 32 & Z. Benkhaldoun, C. Rinner & \\
Wise Observatory, 70 cm & 01:23:01 01:43:59 & 1.0 1.003 & 30.596673 34.762208 883 & 0.700 20 & S. Kaspi & \\
Artemis Observatory & 01:24:04 01:40:12 & 0.3 1.822 & 28.299985 -16.511551 2437 & 1.000 30 & A. Burdanov, J. de Wit & \\
GAL Hassin Robotic Telescope & 01:24:34 01:51:07 & 9.0 10.339 & 37.939377 14.020610 648 & 0.400 18 & A. Nastasi & \\
Serra La Nave (SLN), 91 cm & 01:15:03 01:43:59 & 1.0 1.244 & 37.692467 14.973149 1729 & 0.910 23 & A. Frasca, G. Catanzaro, G. Occhipinti & \\
Teide Observatory, TAR2 & 01:27:38 01:32:14 & 1.0 1.02 & 28.298441 -16.510282 2416 & 0.457$\ $ 6 & M. Alarcón, J. Licandro, M. Serra-Ricart & \\
Cancelada & ... ... & ... ... & 36.461066 -5.054470 76 & ... ... & J. Calvo Fernández & Clouded\\
Observatorio Astronómico de Forcarei & ... ... & ... ... & 42.610608 -8.370880 729 & ... ... & H. González & Clouded\\
Sabadell & ... ... & ... ... & 41.550046 2.090130 263 & ... ... & C. Perelló & Clouded\\
\bottomrule
\end{tabular}
\caption{Details of the observing sites. All the stations that attempted observation are listed.}
\label{tab:observations}
\end{table*}

\begin{landscape}
\section{Pairwise marginal plot for the obtained posterior sample}
    \begin{figure}[ht]
        \centering
        \includegraphics[width=25cm]{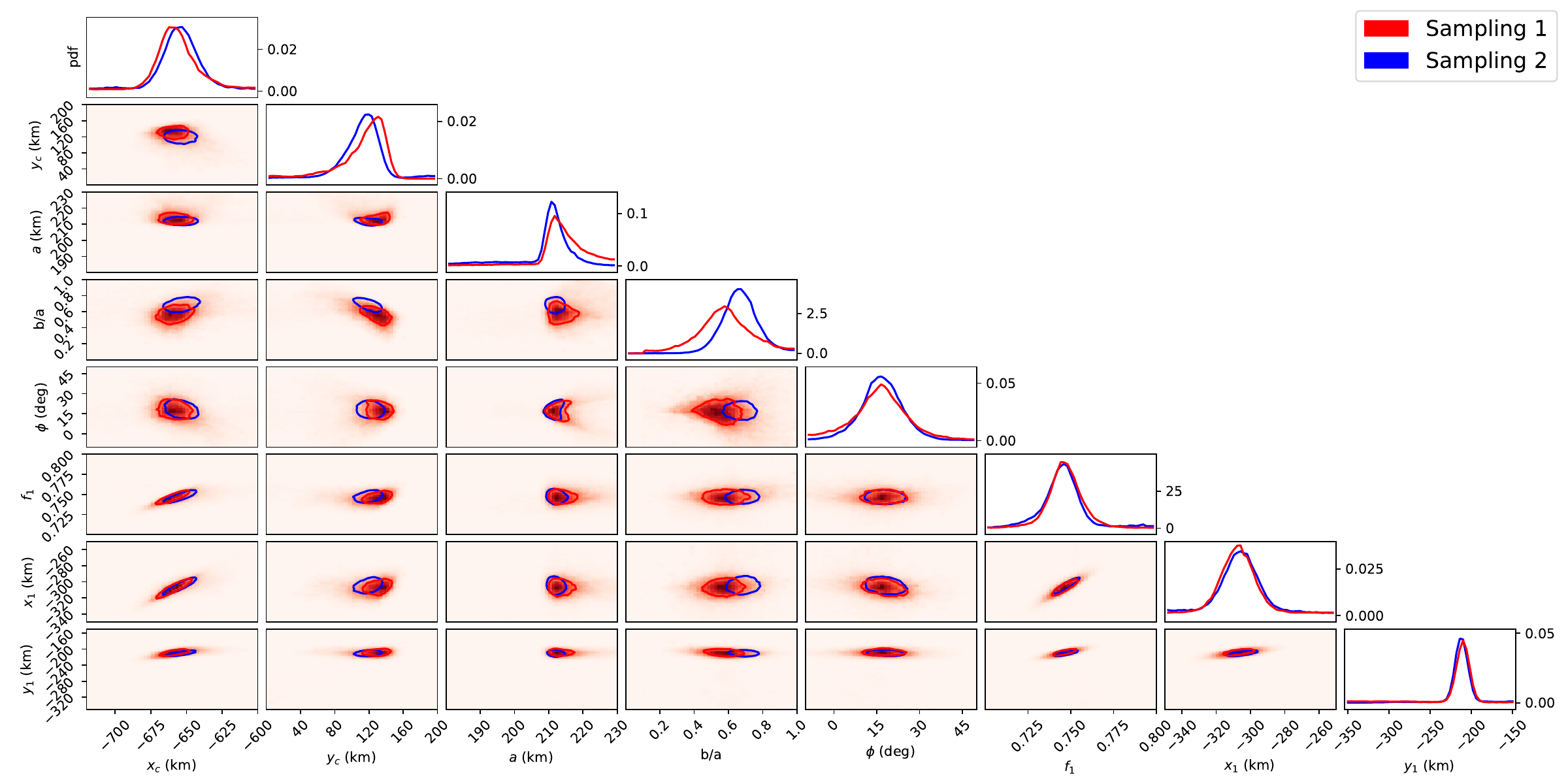}  
        \caption{Pairwise marginal plot for our two MCMC samples of the posterior density. There are no signs of other modes representing alternative solutions. In the bidimensional plots, the represented region encircles 39.4 \% of the samples (representing the $1\sigma$ region). The red shading in these regions is a bidimensional histogram for our preferred sampling 1.}
        \label{fig:cornerplot}
    \end{figure}
\end{landscape}

\end{appendix}

\end{document}